%% file: main.tex
\documentclass[
twocolumn,
english,
aps,
pra,
longbibliography,
superscriptaddress,
amsmath,
amssymb,
floatfix,
]{revtex4-1}
\usepackage{amsthm}
\usepackage{amsfonts}
\usepackage{siunitx}
\usepackage[dvipsnames]{xcolor}
\usepackage{amsmath}
\usepackage{amssymb}
\usepackage{graphicx}
\usepackage{verbatim}
\usepackage[colorlinks]{hyperref}
\usepackage{tikz}
\usepackage{pgfplots}
\usepackage{adjustbox}
\usepackage{braket}
\usepackage{xcolor}
\usepackage{physics}
\usepackage{amssymb} 
\usepackage{graphicx}
\usepackage{dcolumn}
\usepackage{bm}
\usepackage{mathtools}
\usepackage{hyperref}
\usepackage{mathrsfs}
\usepackage{dashrule}
\usepackage{caption}
\usepackage{subcaption}
\usepackage[version=4,arrows=pgf-filled,
textfontname=sffamily,
mathfontname=mathsf]{mhchem}

\usepackage[font=small,labelfont=bf,
   justification=justified,
   format=plain]{caption} 
   
\definecolor{linkcolor}{RGB}{0,83,166}
\hypersetup{
  colorlinks = true,
  allcolors = {linkcolor}
}

\begin{document}

\title{Magnetic Hysteresis Experiments Performed on Quantum Annealers}

\author{Elijah Pelofske}
\email[]{epelofske@lanl.gov}
\affiliation{Information Systems \& Modeling, Los Alamos National Laboratory}

\author{Frank Barrows}
\email[]{fbarrows@lanl.gov}
\affiliation{Theoretical Division, Quantum \& Condensed Matter Physics, Los Alamos National Laboratory}
\affiliation{Center for Nonlinear Studies, Los Alamos National Laboratory}

\author{Pratik Sathe}
\email[]{psathe@lanl.gov}
\affiliation{Theoretical Division, Quantum \& Condensed Matter Physics, Los Alamos National Laboratory}
\affiliation{Information Science and Technology Institute, Los Alamos National Laboratory}

\author{Cristiano Nisoli}
\email[]{cristiano@lanl.gov}
\affiliation{Theoretical Division, Quantum \& Condensed Matter Physics, Los Alamos National Laboratory}
\affiliation{Center for Nonlinear Studies, Los Alamos National Laboratory}
\affiliation{Information Science and Technology Institute, Los Alamos National Laboratory}

\begin{abstract}

\input{main_abstract.tex}

\end{abstract}

\maketitle

\input{main_text}

\clearpage
\newpage

\appendix

\input{appendix_sections}


\bibliographystyle{apsrev4-1}
\bibliography{references}
\end{document}

%% file: main_abstract.tex
While quantum annealers have emerged as versatile and controllable platforms for experimenting on correlated spin systems, the important phenomenology of magnetic memory and hysteresis remain unexplored on hardware designed to escape metastable states via quantum tunneling. Here, we present the first general protocol to experiment on magnetic hysteresis on programmable quantum annealers, and implement it on three D-Wave superconducting qubit quantum annealers, using up to thousands of spins, for both ferromagnetic and disordered Ising models, and across different graph topologies. We observe hysteresis loops whose area depends non-monotonically on quantum fluctuations, exhibiting both expected and unexpected features, such as disorder-induced steps and non-monotonicities. Our work establishes quantum annealers as a platform for probing non-equilibrium emergent magnetic phenomena, thereby broadening the role of analog quantum computers into foundational questions in condensed matter physics.

%% file: main_text.tex
Complex models of frustrated magnets~\cite{King_2021_geometrically_frustrated, Scholl_2021, King_2018, Kairys_2020, harris2018phase, qubit_spin_ice, Ebadi_2021} and topological spin liquids~\cite{semeghini2021probing, qubit_spin_ice, lopez2023kagome} have recently been realized on analog quantum annealers~\cite{PhysRevE.58.5355, Morita_2008, farhi2000quantumcomputationadiabaticevolution}, increasingly used as programmable dynamical laboratories with controllable quantum fluctuations and direct access to individual spin configurations.

Nevertheless, a gap remains. Systematic studies of hysteresis, the landmark feature of magnetic systems, are still lacking, limiting our understanding of out-of-equilibrium, memory-preserving dynamics under external fields. Although some works have explored field-induced phases~\cite{PRXQuantum.2.030317,lopez2023kagome}---notably, one study on out-of-equilibrium magnetization response that inspired the present work~\cite{PRXQuantum.2.030317}---these were context-specific, did not provide a general framework for magnetic hysteresis, and did not demonstrate memory retention through a full cycle of longitudinal field sweeps.
The challenge  stems from the hardware design. Annealing protocols  exploit quantum tunneling to escape local minima for combinatorial optimization \cite{Santoro_2002}, inherently erasing path-dependent memory. This conflicts with hysteresis, which relies on memory retention and metastability. Nonetheless, a controlled activation of the {\em transverse} field has been shown to retain a disordered spin configuration while ``kicking"  topological defects~\cite{qubit_spin_ice,lopez2023kagome,lopez2024quantum}.

\begin{figure*}[ht!]
    \includegraphics[width=0.99\linewidth]{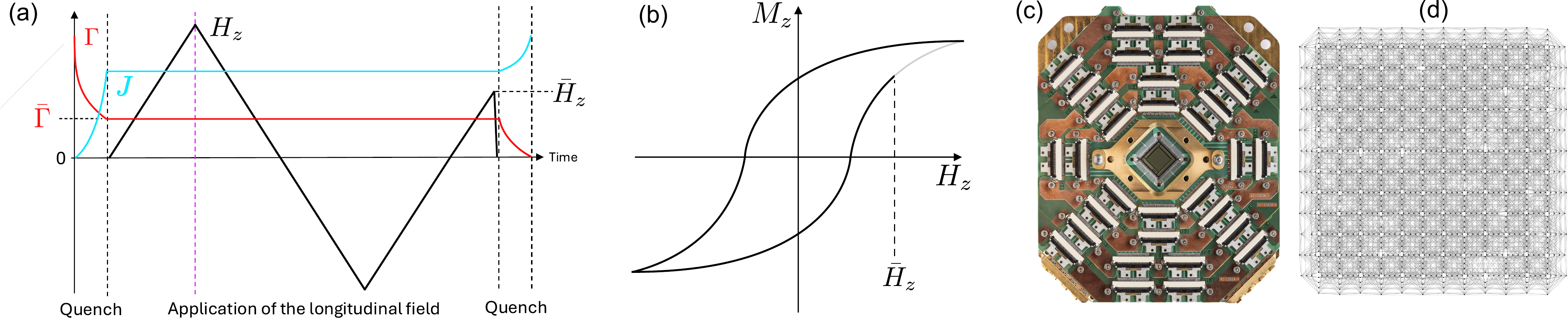}
    \caption{{\bf Schematic Magnetic Hysteresis Protocol and D-Wave QPU.}  (a) Schematic showing the time dependence of $\Gamma$, $J$, and $H_z$ in Eq.~(\ref{equation:Transverse_Ising}), used to generate the magnetization at field $\bar{H}_z$. Initially, the annealing parameter $s$ is rapidly increased to a fixed value, setting $\Gamma$ and $J$ to their target values. $\Gamma$ (red) and $J$ (cyan) remain fixed while the longitudinal full sweep of $H_z$ traces the conceptual schematic hysteresis loop shown in (b), with readout occurring at many intermediate points denoted as $\bar{H}_z$ at simulation times after the dashed vertical purple line. A final quench of $\Gamma$ and $J$ allows final read-out in the $Z$ basis. Initial and final quenches last 0.5 microseconds; field sweeps span up to 10 microseconds (see Supplementary Information~\ref{section:appendix_protocol} for more implementation details). (c) A Pegasus D-Wave QPU chip and (d) the native graph architecture of \texttt{Advantage2\_prototype2.6}, one of three quantum annealing processors used in this study. } 
\label{fig:Figure_1_time_dependent_schedules}
\end{figure*}

We demonstrate for the first time quantum annealers as experimental platforms to performs hysteresis measures on collective spin systems. We apply a time-varying longitudinal field to a system of interacting qubits, while maintaining quantum fluctuations that facilitate transitions and show that the system retains memory: the outcomes are explicitly dependent on the history of the applied field, resulting in hysteresis. Importantly, memory is tunable and varies non-monotonically with the magnitude of the transverse field, which sets the strength of quantum fluctuations.

We showcase our methodology on three representative models across three different machines. In simple high-coordination ferromagnets, we observe the expected box-shaped hysteresis. In systems of disordered coupling we observe Barkhausen noise, also as expected. In lower-coordination, disordered systems, we detect intriguing pinched non-monotonicities which have been previously observed in certain spin glasses, and whose mechanism remains poorly understood. 

Our approach is not merely a simulation but an experiment on superconducting qubits, conceptually identical to techniques used to probe magnetic materials in a laboratory.

\section{Protocol}
\label{section:protocol_high_level}

We employ three D-Wave quantum processing units (QPUs), based on superconducting flux qubits~\cite{johnson2011quantum, Lanting_2014, Bunyk_2014}, interconnected by couplers arranged in machine-specific graphs known as Pegasus~\cite{dattani2019pegasussecondconnectivitygraph, boothby2020nextgenerationtopologydwavequantum} and Zephyr~\cite{zephyr}. The chip ids of three QPU's are: \texttt{Advantage\_system4.1} (5267 qubits), \texttt{Advantage\_system6.4} (5612 qubits), and \texttt{Advantage2\_prototype2.6} (1248 qubits). They implement the  the time-dependent transverse-field Ising Hamiltonian~\cite{de1963collective}
\begin{equation}
    \mathcal{H} = - \Gamma \sum_i \hat{\sigma}_{x}^{i} +  \sum_i h_i \hat{\sigma}_z^{i} + \sum_{\langle ij \rangle} J_{ij} \hat{\sigma}_z^{i} \hat{\sigma}_z^{j},
    \label{equation:Transverse_Ising}
\end{equation}
where $\hat{\sigma}^i$ are Pauli matrices at node $i$, and $J_{ij}$ are couplings along edges ${\langle ij \rangle}$. 

When $\Gamma = 0$, the ``longitudinal" Hamiltonian is diagonal in the $\hat{\sigma}_z$ basis. For $\Gamma \neq 0$, the transverse term does not commute with $\hat{\sigma}_z$, introducing quantum fluctuations that drive  transitions among states of the longitudinal Hamiltonian. An Ising spin model is programmed on D-Wave hardware by specifying the interaction values $J_{ij}$ for each coupler, and $h_i$ for each qubit. 

In standard quantum annealing, the energy scale $J_{ij}$ increases while $\Gamma$ gradually reduces to zero, driving the system from a quantum paramagnetic state toward a ``classical" configuration by progressively suppressing quantum fluctuations. Specifically, in D-Wave quantum annealer the Hamiltonian above takes the form 
\begin{align}
    \begin{split}
        {\mathcal H} =& - \frac{A(s)}{2}  \sum_i \hat\sigma_x^{i} \\
        &+ \frac{B(s)} {2} \left( g(t) \sum_i \tilde h_i \hat\sigma_z^{i} + \sum_{\langle ij \rangle} \tilde J_{i j} \hat\sigma_z^{i} \hat\sigma_z^{j} \right). 
    \end{split}
     \label{equation:QA_Hamiltonian_h_gain}
\end{align}
$A(s)$ and $B(s)$ are hardware-specific functions, decreasing and  increasing respectively in the annealing parameter $s$ which varies in time from 0 to 1 (see Supplementary Information~\ref{appendix:DWave_calibrated_schedules}). Critically, the global time-dependent multiplier $g(t)$ (or ``h-gain'') scales all local fields and can be varied independently from $s(t)$: this hardware feature is central to our experiments.
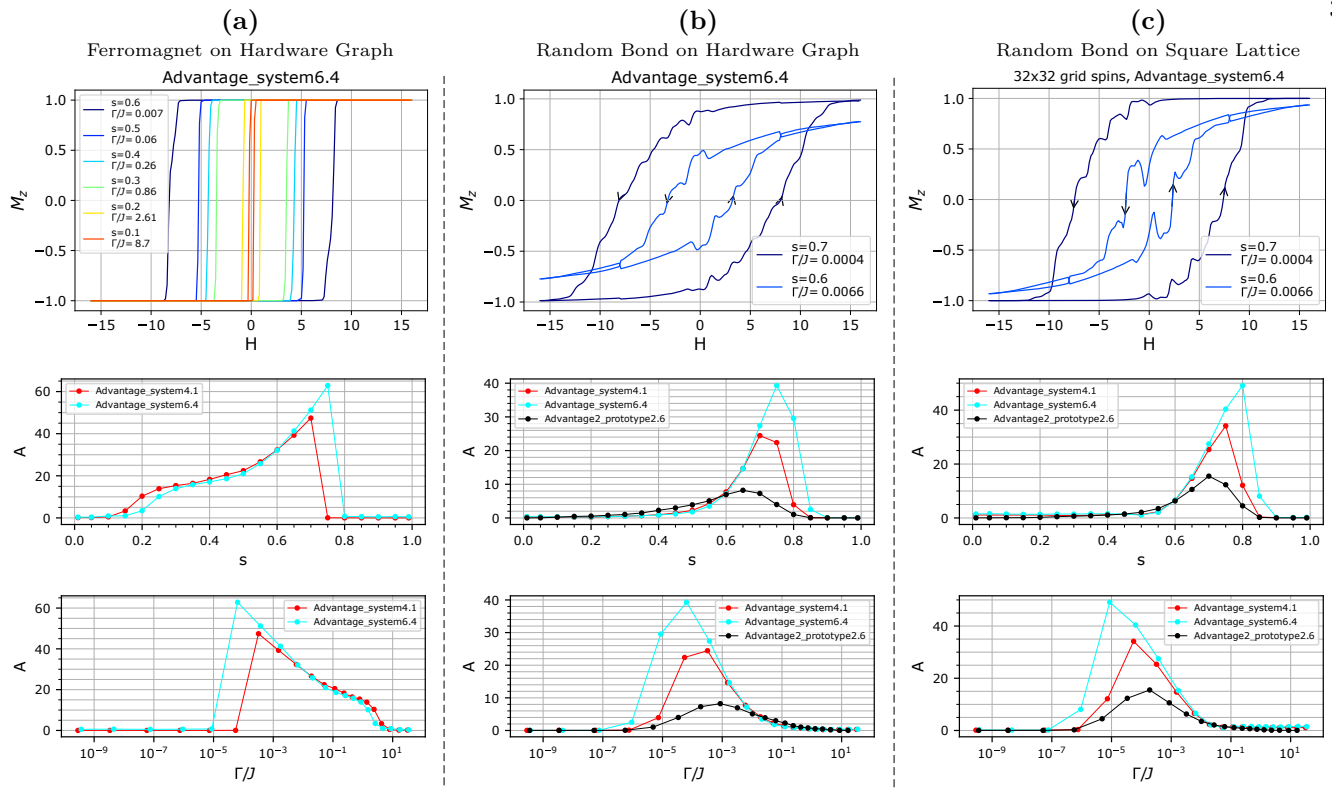
\begin{figure*}[ht!]
    \begin{picture}(0,0)
         \put(-220,1){\scriptsize \shortstack{\normalsize{\textbf{(a)}} \\ Ferromagnet on Hardware Graph}}
         \put(-50,1){\scriptsize \shortstack{\normalsize{\textbf{(b)}} \\ Random Bond on Hardware Graph}}
         \put(124,1){\scriptsize \shortstack{\normalsize{\textbf{(c)}} \\ Random Bond on Square Lattice}}
    \end{picture}
    \includegraphics[width=0.999\linewidth]{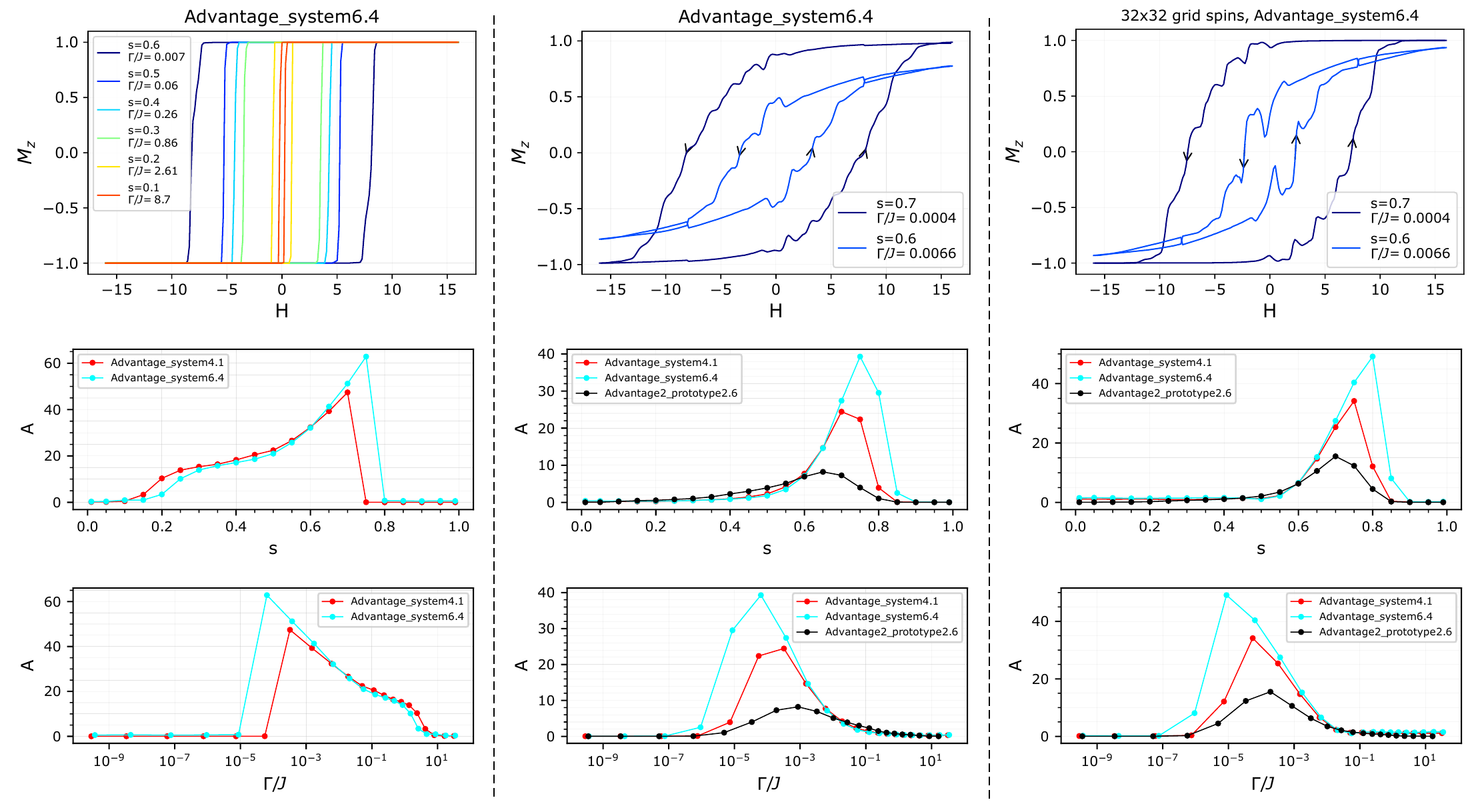}
    \caption{{\bf Magnetic Hysteresis Experimental Results on D-Wave QPUs.} The three columns corresponds to: (a) {\it Ferromagnet on the Hardware Graph} (b) {\it Binary Random Bond ($\pm J$) Model on the Hardware Graph}, (c) {\it Binary Random Bond ($\pm J$) Model on a Square Lattice (with open boundary conditions)}. Top row: plots of normalized magnetization $M_z$,vs. longitudinal field $H$, for the three models, all implemented on the same \texttt{Advantage\_system6.4} processor (see Fig.~\ref{fig:Figure_3_hardware_defined_spin_glass_hysteresis} and Fig.~\ref{fig:Figure_4_2D_hysteresis_curves} for results across all three devices) exhibit closed-loop magnetic hysteresis. The arrows along the magnetization curves denote simultaneously the direction of the longitudinal field sweep and the time progression of the protocol. The ferromagnet loops exhibit a ``boxed'' shape indicative of high coordination and mean-field behavior; the $\pm J$ models show crackling noise and nonmonotonic features expected from structural disorder. Middle row: hysteresis loop area vs. hardware annealing parameter $s$ for three different D-Wave QPUs. Bottom row: hysteresis loop area as a function of equivalent physical ratio $\Gamma/J$ to the $s$ value plots shown in the middle row. }
    \label{fig:Figure_2_hardware_3D_FM_area}
\end{figure*}
Our protocol, illustrated in Fig.~\ref{fig:Figure_1_time_dependent_schedules}, begins by rapidly quenching the annealing parameter $s$ from 0 to a target value in the range $s = 0$–$1$, within approximately $0.5~\mu$s, setting the transverse field $\Gamma$ and the coupling strength $J$. Holding $s$ (and thus $J$ and $\Gamma$) fixed in time we apply a uniform longitudinal field $h_i = H_z$ through various values of the loop, by programming the global scaling function $g(t)$. The sweeps, possibly positive and negative, stop at a field $\bar{H}_z$, after which $g(t)$ is quickly brought to zero and $s$ is ramped to 1, quenching $\Gamma$ before readout of qubits are then measured in the $\hat{\sigma}_z$ basis, from which the average magnetization $M_z$, as well as other observables, can be extracted by repeated hardware anneal-readout cycles (in this study, we measure $2{,}000$ samples for each $\bar{H}_z$). This cycle is repeated across a range of $\bar{H}_z$ values to reconstruct the full probe–response dynamics. 

Each Ising spin maps to a physical qubit on the hardware lattice. Auto-scaling of programmed Ising model coefficients is disabled to preserve intended energy scales. The D-Wave QPUs operate at $\sim$15~mK, and experiment timescales exceed closed quantum system decoherence times~\cite{king2022coherent, king2024computationalsupremacyquantumsimulation, king2023quantum, tindall2025dynamicsdisorderedquantumsystems, PhysRevResearch.2.033369}, placing the dynamics in a quasistatic, noisy, open-system regime~\cite{Amin_2015, PhysRevApplied.19.034053, PhysRevApplied.8.064025}. All experiments in this study use a total simulation time, e.g., annealing time, of $11.2 \mu$ seconds (see Supplementary Information~\ref{section:appendix_protocol}).

\section{Experiments}
\label{section:experiments}

We test our hysteresis protocol on three distinct types of Ising models, and demonstrate hysteresis and memory. 

We begin with the ferromagnet on the full hardware graph, a natural test case. Fig.~\ref{fig:Figure_2_hardware_3D_FM_area}(a) shows average magnetization per spin versus longitudinal field on \texttt{Advantage\_system6.4}, revealing clear hysteresis and memory. The rectangular shape is consistent with the mean-field behavior expected in ferromagnets of high coordination. Results for \texttt{Advantage\_system4.1}, nearly identical, are shown in Supplementary Information~\ref{section:appendix_remaining_ferromagnetic_model_hysteresis}.

Our second and third models exhibit considerably more complex behavior. We realize a Binary Random Bond Model on the full hardware graph and also on a Square Lattice with open boundary conditions. In the latter case we embed the square lattice using a fixed hardware-native subgraph isomorphism using the Glasgow Subgraph Solver~\cite{mccreesh2020glasgow}, embedding square grids of size $32 \times 32$ on \texttt{Advantage\_system4.1} and \texttt{Advantage\_system6.4}, and size $26 \times 26$ on \texttt{Advantage2\_prototype2.6}. 
In both cases, all couplings have fixed magnitude $J$ but random sign ($\pm J$) in equal ratio, a coexistence of ferromagnetic and antiferromagnetic interactions known to lead to frustration and glassy behavior~\cite{ozeki1987phase,toulouse1987theory}. 
While full spin glass characterization is beyond the scope of the current study, here we highlight the potential of our method by showing that the introduction of disorder leads to considerably more complex hysteresis. In particular, it should give rise to Barkhausen jumps and crackling noise, which are hallmarks of disordered magnetic systems~\cite{barkhausen1919zwei,perkovic1995avalanches,sethna2001crackling,puppin2000barkhausen,callegaro2003barkhausen}. 

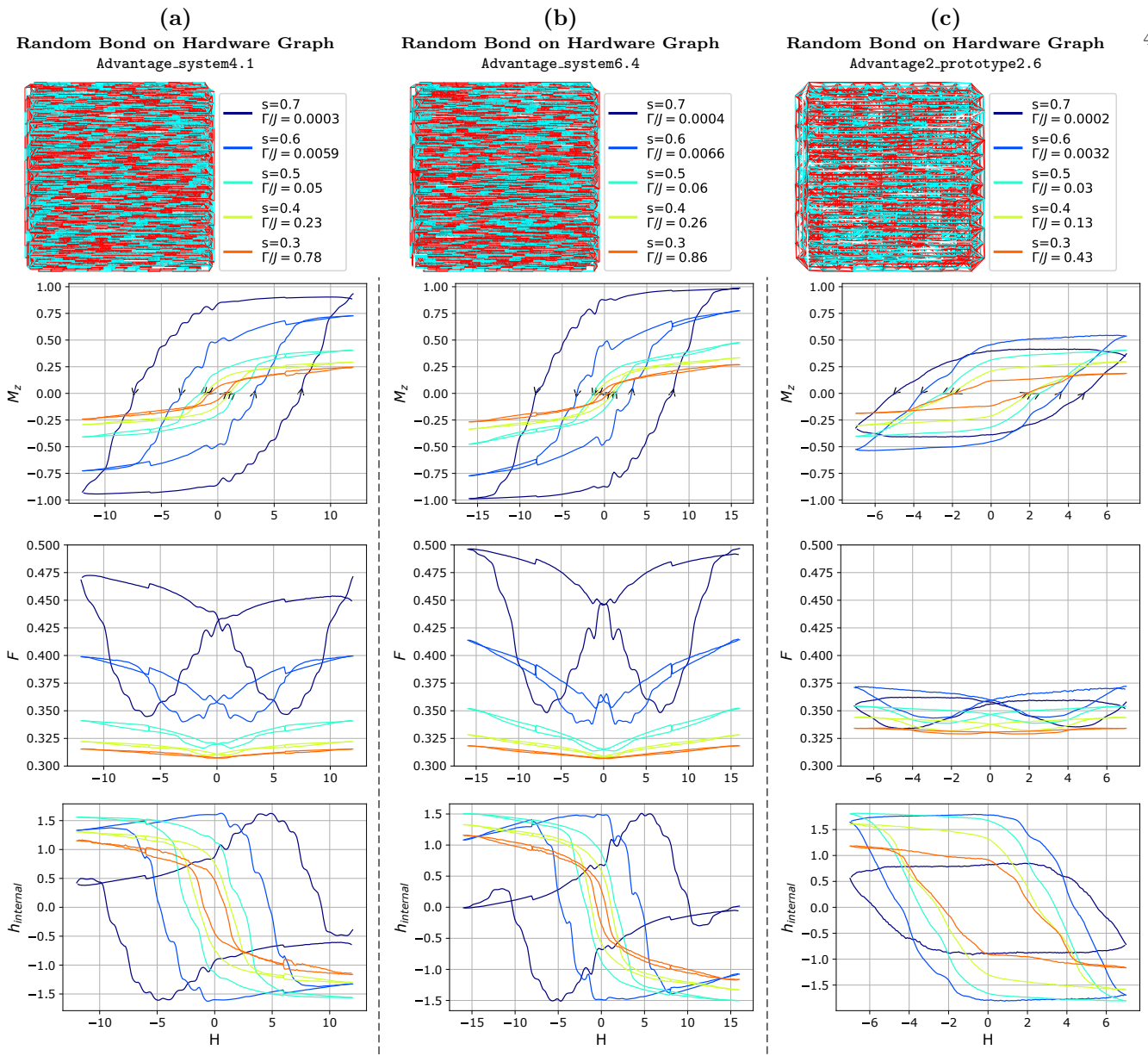
\begin{figure*}[ht!]
     \centering
      \begin{picture}(0,0)
         \put(10,90){\scriptsize \shortstack{\normalsize{\textbf{(a)}} \\ \textbf{Random Bond on Hardware Graph} \\ \texttt{Advantage\_system4.1}}}
    \end{picture}
    \includegraphics[width=0.32\linewidth]{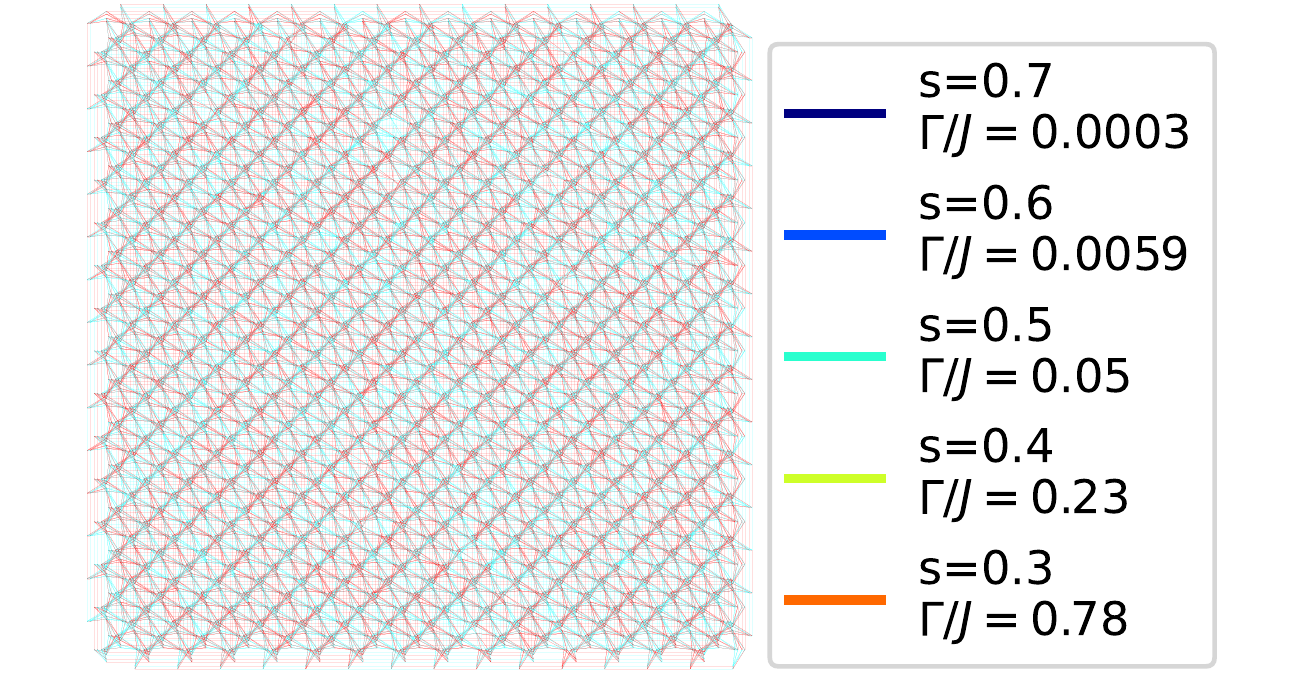}
    \begin{picture}(0,0)
         \put(10,90){\scriptsize \shortstack{\normalsize{\textbf{(b)}} \\ \textbf{Random Bond on Hardware Graph} \\ \texttt{Advantage\_system6.4}}}
    \end{picture}
    \includegraphics[width=0.32\linewidth]{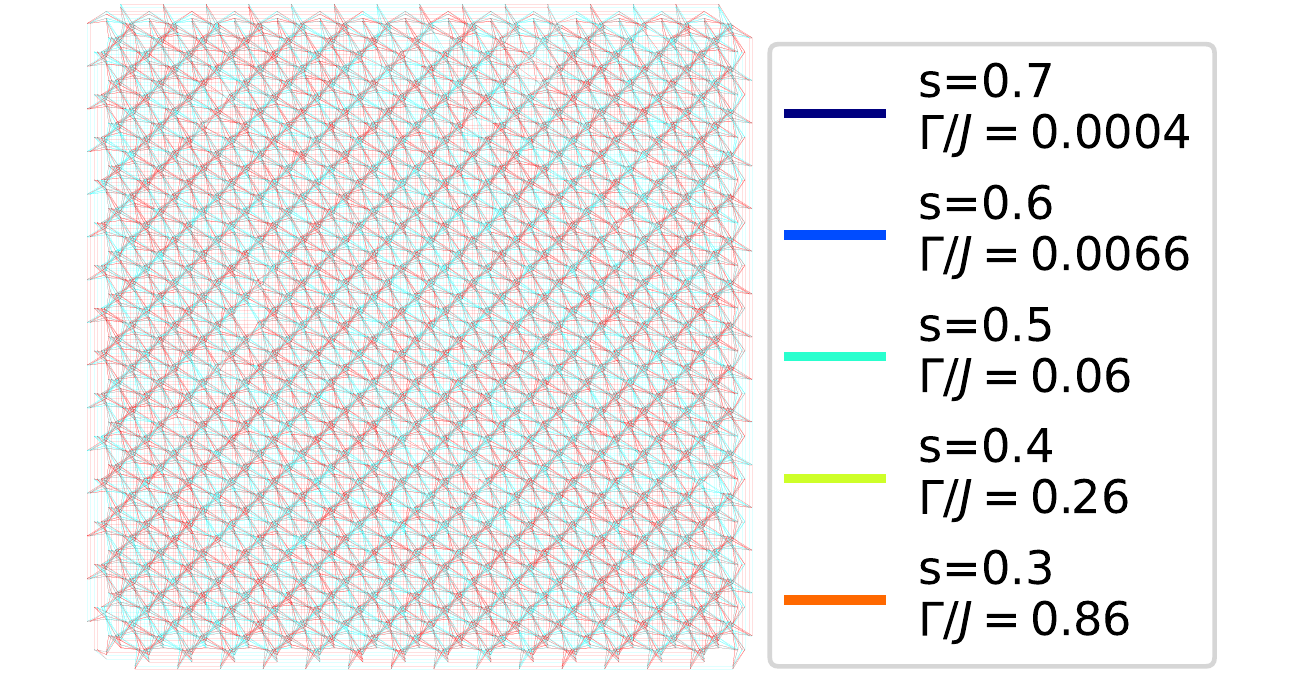}
    \begin{picture}(0,0)
         \put(10,90){\scriptsize \shortstack{\normalsize{\textbf{(c)}} \\ \textbf{Random Bond on Hardware Graph} \\ \texttt{Advantage2\_prototype2.6}}}
    \end{picture}
    \includegraphics[width=0.32\linewidth]{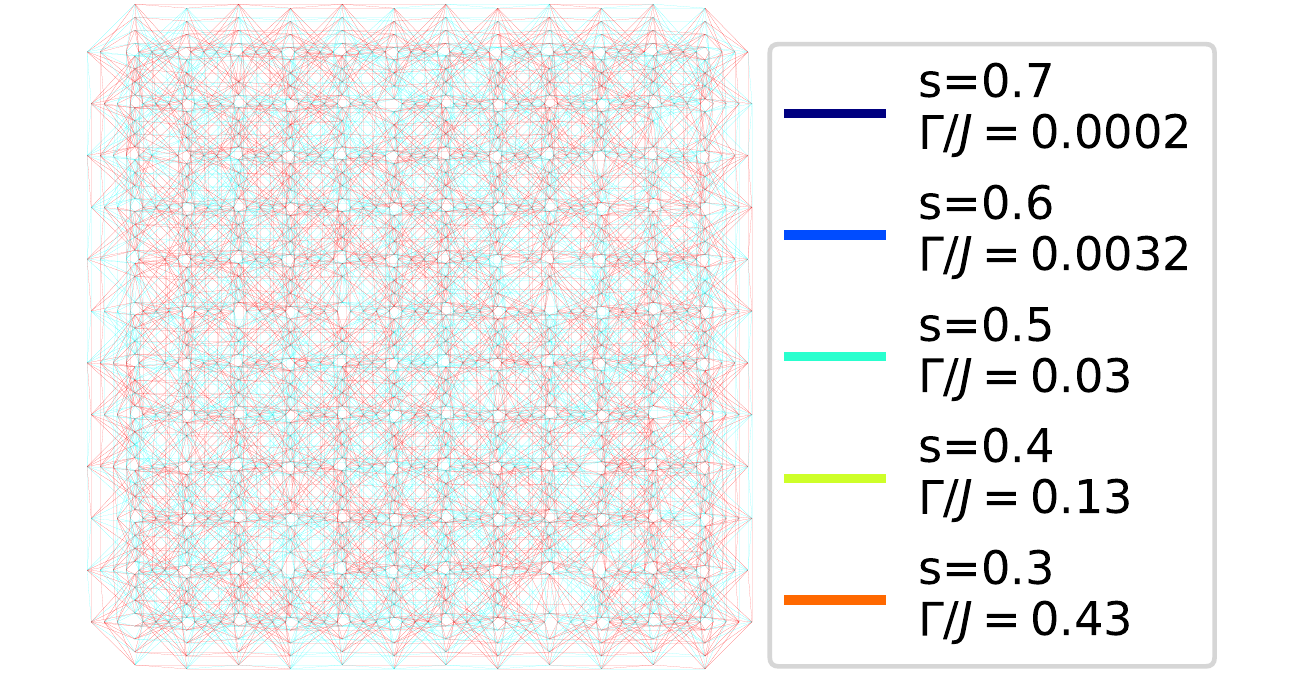}
`    \includegraphics[width=0.999\linewidth]{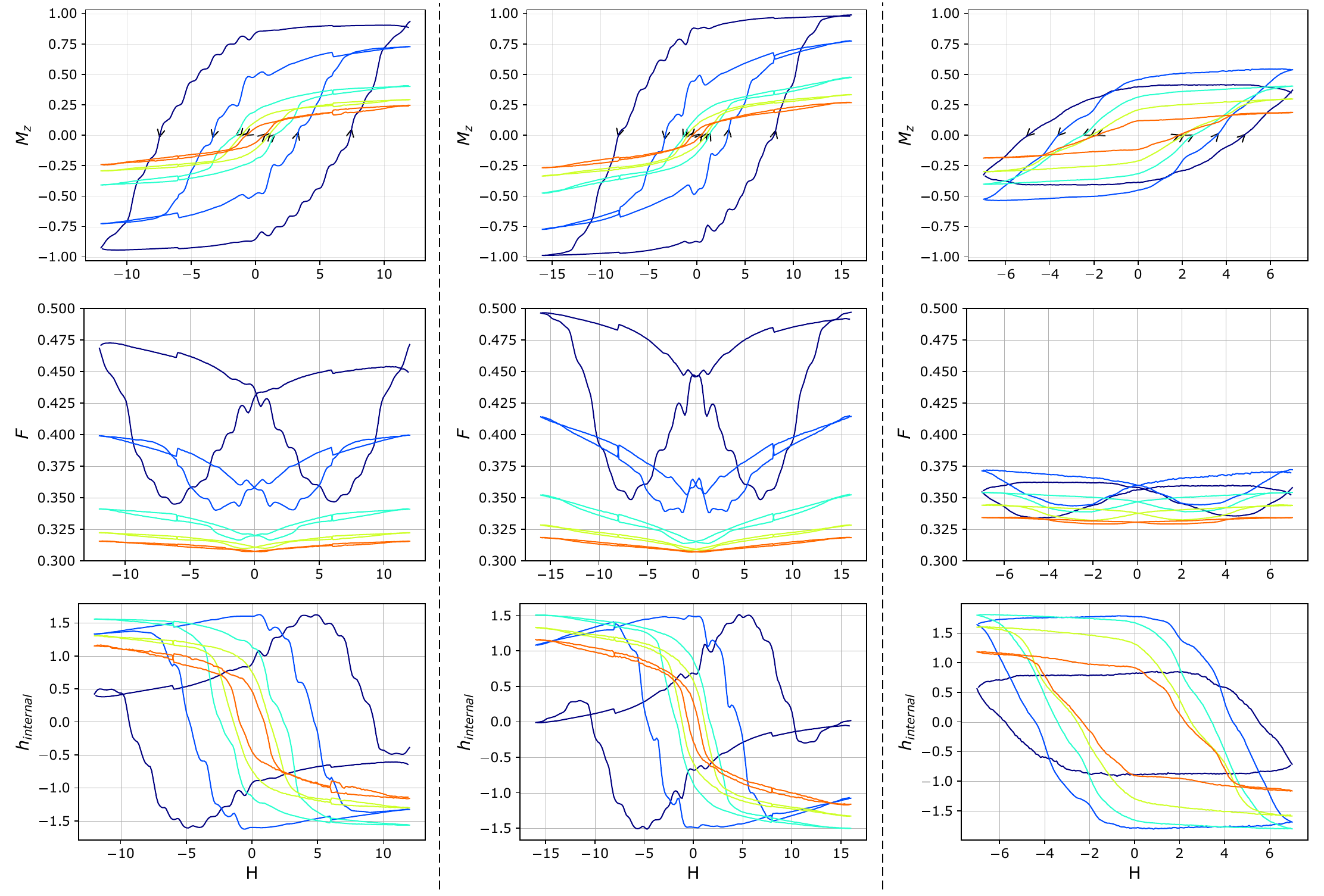}
    \caption{{\bf Magnetic Hysteresis for the $\pm J$ Model on Full Hardware Graphs.} The three distinct hardware-defined $\pm J$ models are shown on top ($J=-1$ edge coupling is cyan, and $J=+1$  is red). In the following rows, we present average magnetization $M_z$, fraction of frustrated bonds ${\cal F}$, and internal field $h^{\mathrm{internal}}$, plotted against the applied field for various values of $s$ (and thus different $\Gamma/J$ ratios) specified in the legends of the top row. For sufficiently large $\Gamma/J$, the hysteresis loop area almost disappears as the model approaches a quantum paramagnet. The arrows along the magnetization curves simultaneously denote the direction of the longitudinal field sweep and the time progression of the protocol. }
    \label{fig:Figure_3_hardware_defined_spin_glass_hysteresis}
\end{figure*}

\begin{figure*}[th!]
    \centering
      \begin{picture}(0,0)
         \put(-236,2){\scriptsize \shortstack{\normalsize{\textbf{(a)}} \\ \textbf{Random Bond on Square Lattice} \\ \texttt{Advantage\_system4.1}}}
         \put(-60,2){\scriptsize \shortstack{\normalsize{\textbf{(b)}} \\ \textbf{Random Bond on Square Lattice} \\ \texttt{Advantage\_system6.4}}}
         \put(104,2){\scriptsize \shortstack{\normalsize{\textbf{(c)}} \\ \textbf{Random Bond on Square Lattice} \\ \texttt{Advantage2\_prototype2.6}}}
    \end{picture}
    \includegraphics[width=0.999\linewidth]{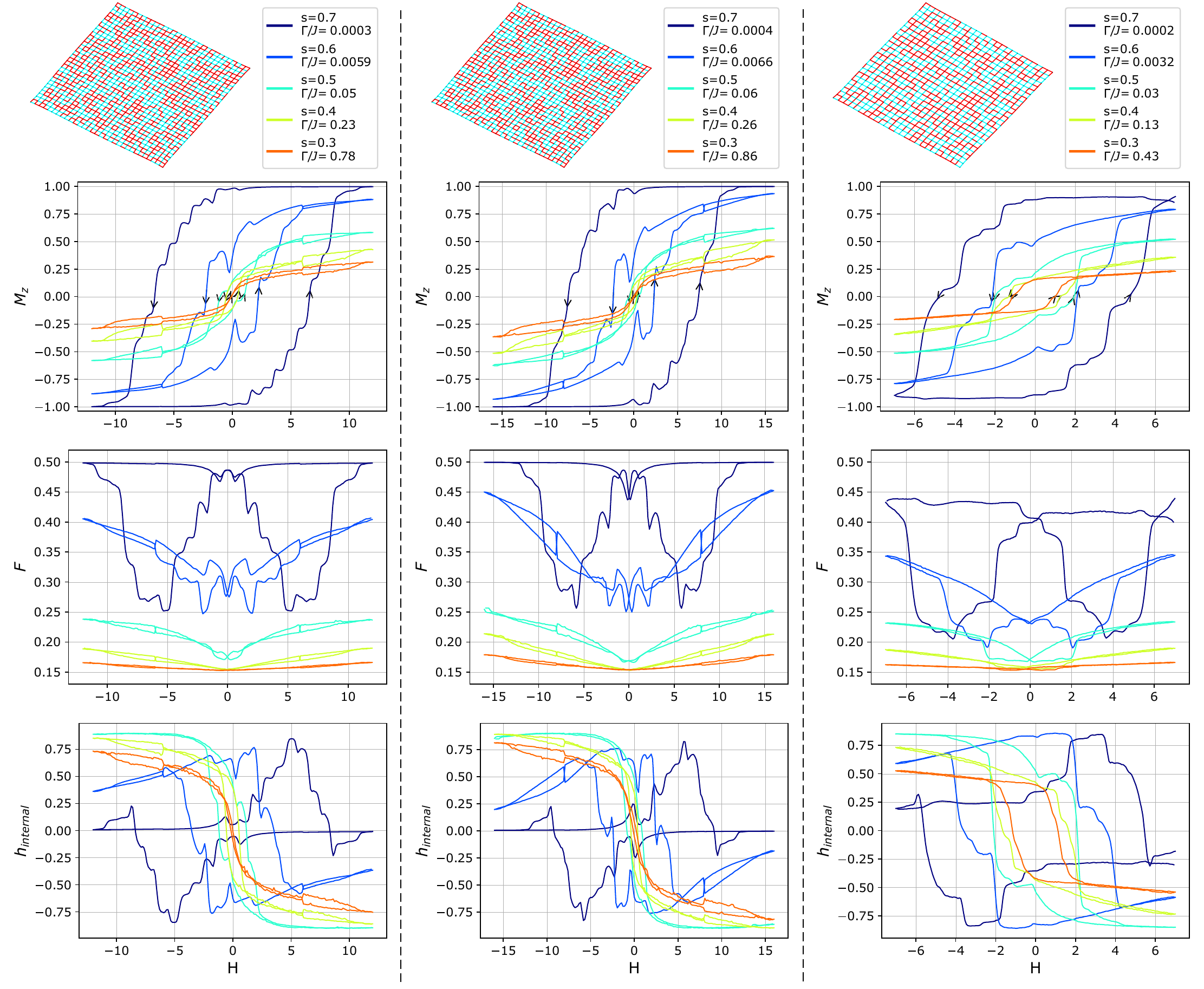}
    \caption{{\bf Magnetic Hysteresis for the $\pm J$ Model Defined on 2D square lattices}. The three distinct $\pm J$ model on a square graph, with open boundary conditions, are shown on top row ($J=-1$ is red, $J=1$ is cyan), corresponding to the three different D-Wave processors. The same $32\times 32$ instance is run on both \texttt{Advantage\_system6.4} (b) and \texttt{Advantage\_system4.1} (a), and a $26\times 26$ $\pm J$ model is run on \texttt{Advantage2\_prototype2.6} (c). In the following rows, we present magnetization $M_z$, fraction of frustrated bonds ${\cal F}$, and internal field $h^{\mathrm{internal}}$, plotted against the applied field for various values of $s$ (and thus different $\Gamma/J$ ratios) specified in the legends of the top row. For sufficiently large $\Gamma/J$, the hysteresis area almost disappears as the model approaches a quantum paramagnet. Unlike in Fig.~\ref{fig:Figure_3_hardware_defined_spin_glass_hysteresis}, the lower coordination of the Ising model allows for full saturation, where as predicted ${\cal F}=0.5$ and $h^{\mathrm{internal}}=0$ in column (a) and (b), but not in (c) due to the lower maximum achievable longitudinal field on the \texttt{Advantage2\_prototype2.6} processor. }
    \label{fig:Figure_4_2D_hysteresis_curves}
\end{figure*}

Fig.~\ref{fig:Figure_2_hardware_3D_FM_area}(b) and (c) report considerably more complex hysteresis loops for these two random bond cases. We observe the anticipated corrugations, suggestive of discrete avalanche-like spin reconfigurations, typical of the disorder-induced Barkhausen effect. We also observe intriguing non-monotonicities in the curves, particularly visible for the square lattice in panel (c), where strong dips emerge. This  phenomenon has been previously reported, e.g., in Ge$_{0.87}$Mn$_{0.13}$Te spin glasses~\cite{krempasky2023efficient,yoshimi2018current}, where is not yet understood, as well as in the competition between internal and collective magnetization of nanolements~\cite{ostman2014hysteresis}. We can speculate that in our $\pm J$ model, competition between ferro- and antiferromagnetic couplings can induce local spin rearrangements, such as reentrant ferromagnetic states. Importantly, the effect is enhanced at intermediate transverse fields, where quantum fluctuations reduce the area of hysteresis but amplify dips and corrugations (see Fig.~\ref{fig:Figure_3_hardware_defined_spin_glass_hysteresis}, Fig.~\ref{fig:Figure_4_2D_hysteresis_curves}), suggesting that tunneling between nearly degenerate states  of different magnetization plays a role.

The memory content can be quantified as the area enclosed by the hysteresis loop and is expected to vary non-monotonically with the ratio $\Gamma/J$: at high $\Gamma/J$, quantum fluctuations should suppress memory as the system approaches quantum paramagnetism; instead, at very low $\Gamma/J$, spins should remain unresponsive. 
Figure~\ref{fig:Figure_2_hardware_3D_FM_area} shows the hysteresis area as a function of the annealing parameter $s$ and the corresponding $\Gamma/J$ ratio across different machines, confirming this non-monotonic behavior. In the ferromagnetic case, the profiles are nearly identical across the two devices, despite slight differences in the hardware graphs, reflecting the  mean-field behavior already mentioned. In contrast, the random bond models show significant variation in hysteresis area at small $\Gamma/J$. This is expected in the full-graph case, because the hardware graphs differ across machines and also because different machines are more or less able to reach full polarization. Machine-intrinsic timescale variations may also affect the hysteresis loop area.

\begin{figure*}[th!]
    \begin{picture}(0,0)
        \put(-12,190){\textbf{(a)}}
    \end{picture}
    \includegraphics[width=0.38\linewidth]{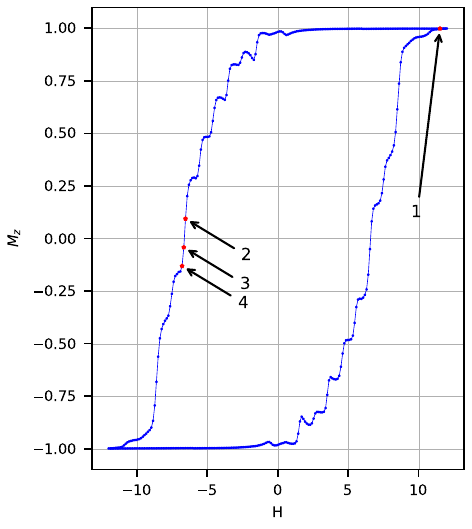}
    \hspace{0.8cm}
    \includegraphics[width=0.55\linewidth]{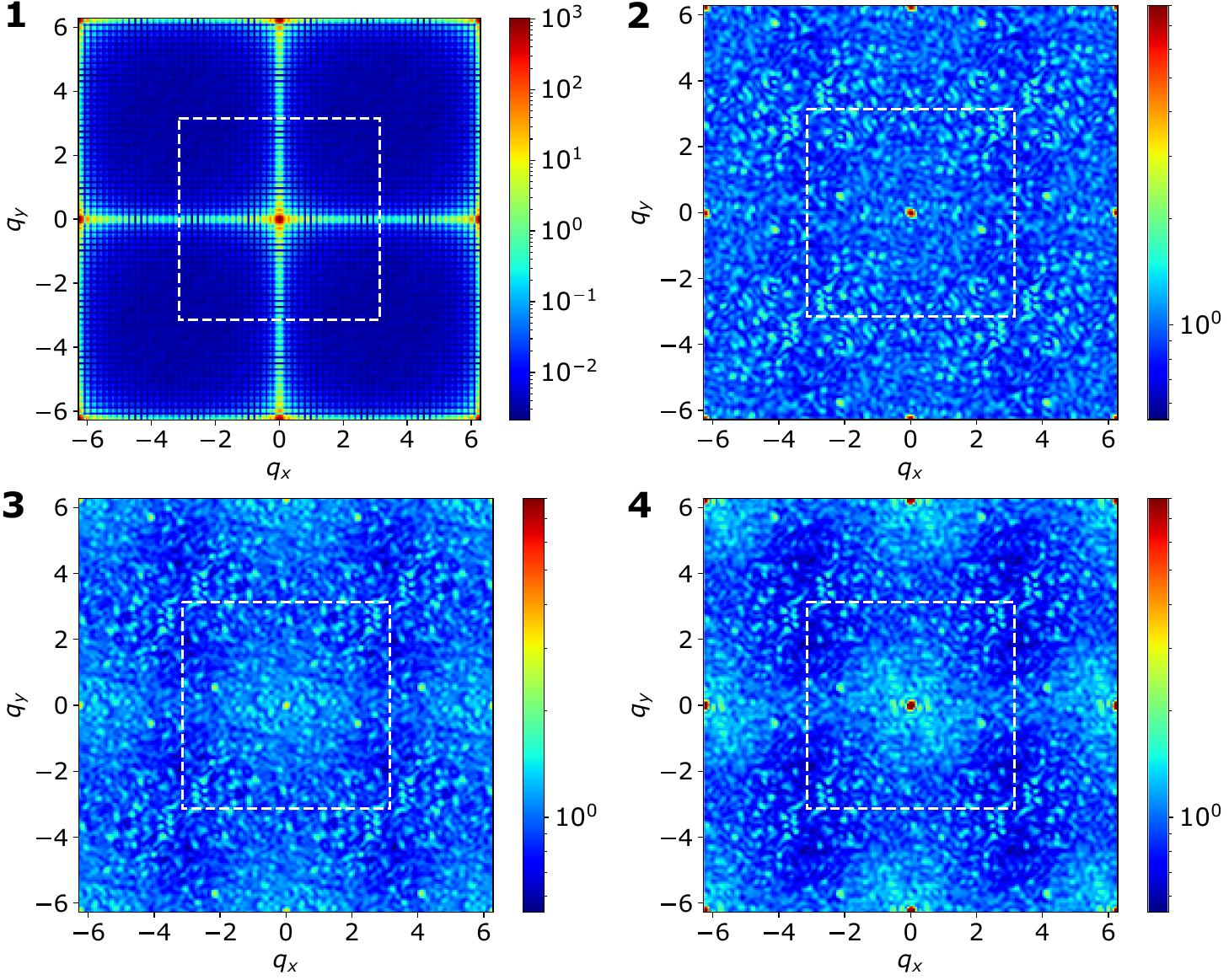}
    \vspace{-0.04cm}
    \tikz{\draw[dashed, thick] (0,0) -- (18.2,0);}
    \vspace{0.02cm}
    \begin{picture}(0,0)
        \put(-12,190){\textbf{(b)}}
    \end{picture}
    \includegraphics[width=0.38\linewidth]{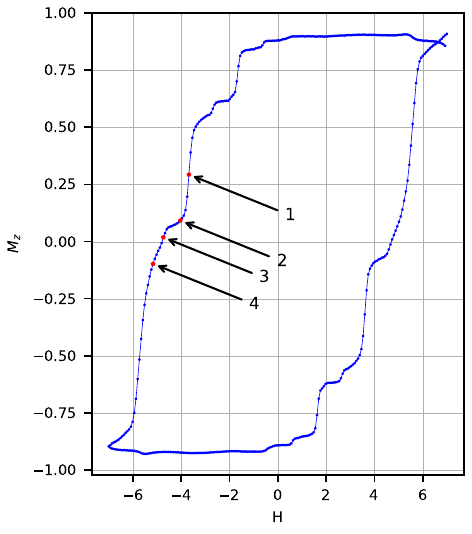}
    \hspace{0.8cm}
    \includegraphics[width=0.55\linewidth]{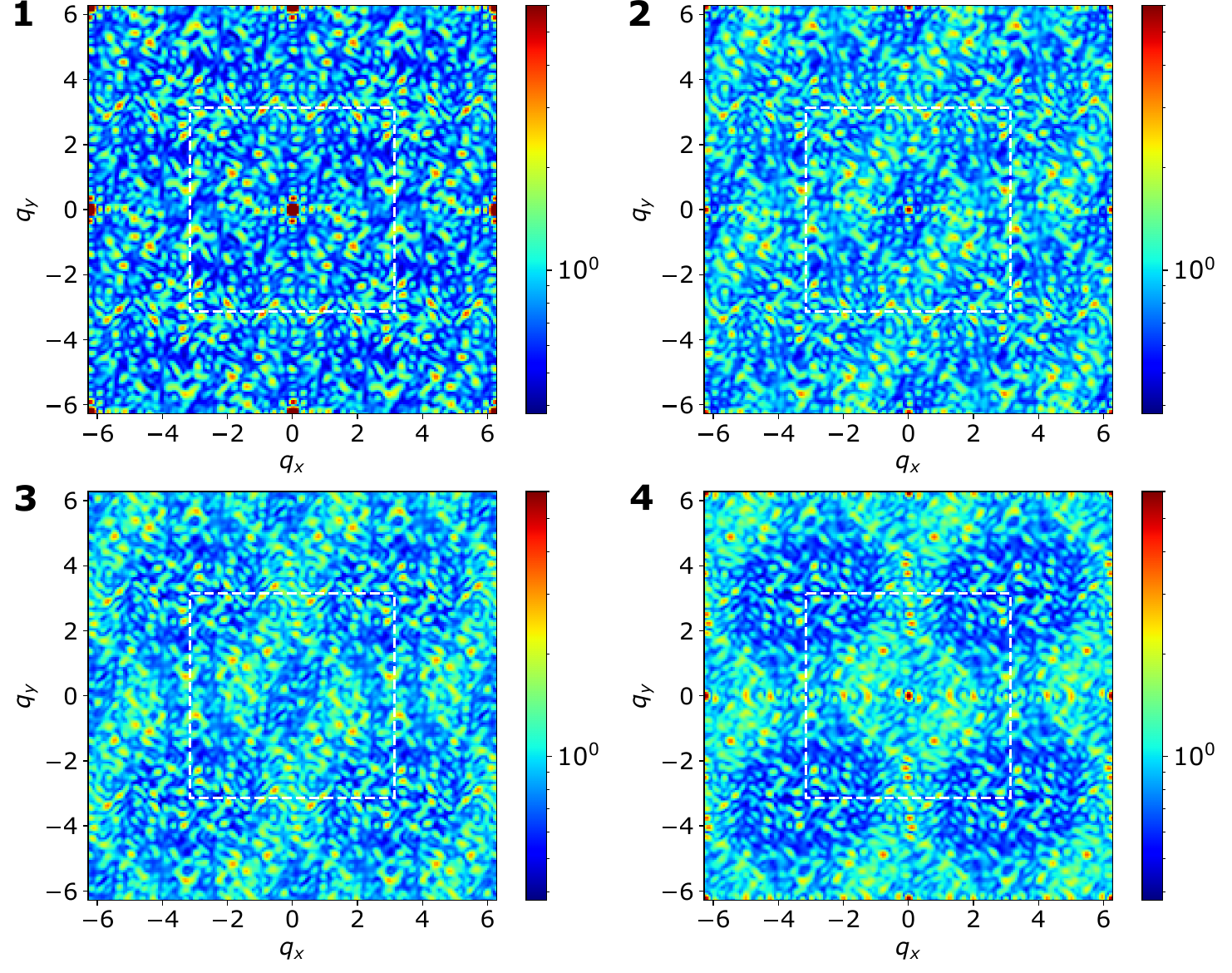}
    \vspace{-0.02cm}
    \tikz{\draw[dashed, thick] (0,0) -- (18.2,0);}
    \vspace{-0.04cm}
    \begin{subfigure}[b]{0.60\linewidth}
    \centering
    \begin{picture}(0,0)
        \put(-165,-14){\textbf{(c)}}
    \end{picture}
    \includegraphics[width=1.0\linewidth]{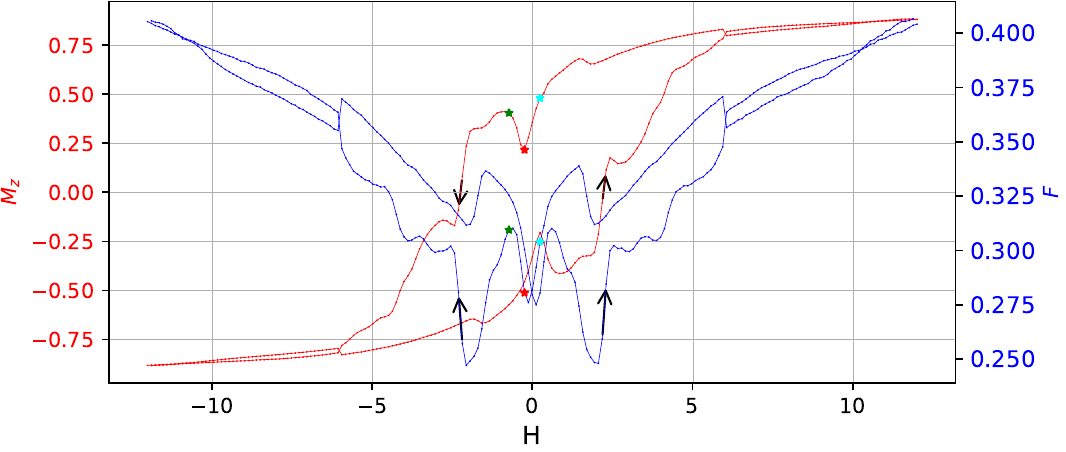}\\
    \end{subfigure}
    \hfill
    \begin{subfigure}[b]{0.39\linewidth}
        \centering
        \includegraphics[width=\linewidth]{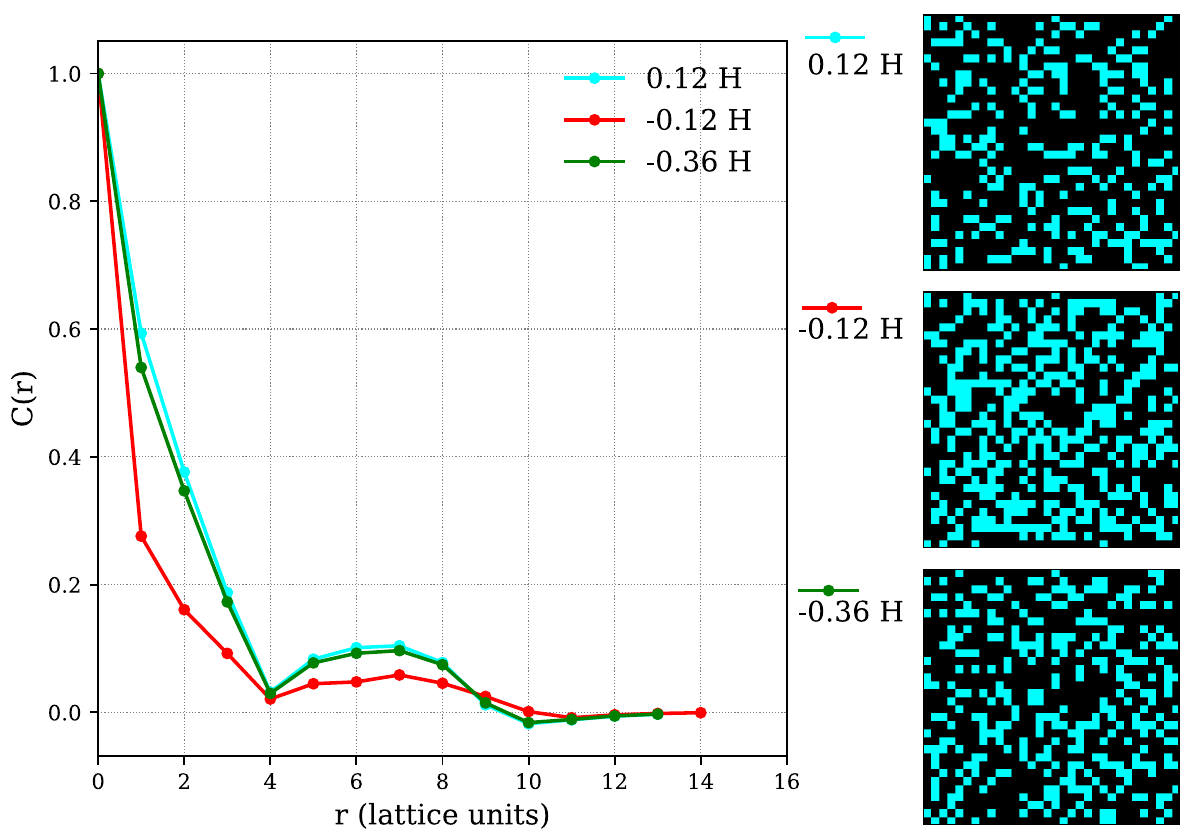}
    \end{subfigure}
    \caption{\textbf{Magnetic Structure Factors and Real Space Configurations.} Hysteresis curves at $s=0.7$ and selected MSFs for (a) the $32\times 32$ $\pm J$ run on \texttt{Advantage\_system4.1} and (b) the $26\times 26$ $\pm J$ run on \texttt{Advantage2\_prototype2.6}. All MSF plots use a log scale $\abs{S(q)}$ heatmap, and the dashed white box outlines the first Brillouin zone. Panel (c) highlights an non-monotonic dip in magnetization by showing $M_z$ and coupler frustration hysteresis ${\cal F}$ data overlayed, from a $32\times 32$ $\pm J$ model run on \texttt{Advantage\_system4.1} at $s=0.6$. Panel (c) also reports correlations along principal axes (line colors correspond to the annotated colors along the hysteresis loop), and example single-sample real space spin configurations, with black (cyan) pixels corresponding to $-1$ ($+1$) spins, from before, at, and after, the non-monotonic magnetization dip. 
    }
    \label{fig:Figure_5_spin_structure_factor}
\end{figure*}
A more complete picture is afforded by Figs.~\ref{fig:Figure_3_hardware_defined_spin_glass_hysteresis} and~\ref{fig:Figure_4_2D_hysteresis_curves}, reporting data for  the Binary Random Bond models, implemented respectively on the full hardware graph and on the square lattice embedding, across the three quantum annealing devices. 

For the full hardware graph case, full magnetic saturation is not reached (Fig.~\ref{fig:Figure_3_hardware_defined_spin_glass_hysteresis}, second row), though \texttt{Advantage\_system6.4} approaches it closely due to its stronger $h$-gain capability (see Supplementary Information~\ref{section:appendix_protocol}). On \texttt{Advantage2\_prototype2.6}, high average connectivity and limited maximum longitudinal field strength prevent $M_z$ from exceeding $0.5$, resulting in minor hysteresis loops, which are generally characterized by smoother profiles because of more localized magnetic responses~\cite{baldwin1973barkhausen,sabhapandit2002hysteresis}.

In the square lattice case (Fig.~\ref{fig:Figure_4_2D_hysteresis_curves}, second row), full magnetization is achieved on both \texttt{Advantage\_system4.1} and \texttt{Advantage\_system6.4}. Interestingly, even at relatively high transverse-to-coupling ratios ($\Gamma/J = 0.78$ and $0.86$, respectively), the system never really approaches the quantum paramagnetic regime, and an intriguing pinched hysteresis appears, more typical of memristors. While vanishing magnetization at zero field is expected under strong quantum fluctuations, the persistence of some history-dependent behavior in this regime highlights the robustness of the memory effect, even as quantum tunneling dominates.
Here too, on \texttt{Advantage2\_prototype2.6}, magnetization never fully saturates, resulting in  minor loops.

We stress here the importance of enabling systematic exploration of minor loops. These are inaccessible to previous ad hoc approaches~\cite{King_2018}, yet essential to enabling experimental studies of return-point memory, training effects, and the emergence of limit cycles in driven spin systems, as we will report in future work.

\section{Microscopic Vistas}
\label{section:microscopic_quantities}

The results presented so far resemble those obtainable in a laboratory setting, but quantum annealers offer unique access to individual spin-level resolution for the extraction of microscopic quantifiers. We highlight a few examples to illustrate the broader potential of our methodology.
For a useful quantifier we can define the average frustration ${\cal F}$, as the fraction of frustrated couplings, or:
\begin{equation}
{\cal F} = \left\langle \frac{1}{2 N_e} \sum_{\langle ij \rangle} \left[1 + \mathrm{sgn}(J_{ij}) \sigma^i_z \sigma^j_z \right] \right\rangle,
\end{equation}
where $\mathrm{sgn}$ is the sign function, $\langle \cdot \rangle$ denotes a sampling average, and $N_e$ is the total number of couplers, and the sum is over all  qubits. 

We can also extract internal ``forces'', such as the effective internal fields acting on each spin:
\begin{equation}
h^{\mathrm{internal}}_{i} = -\sum_{j \in \partial_i} J_{ij} \sigma^{j}_z,
\label{Equation:internal_field}
\end{equation}
where $\partial_i$ denotes the nodes coupled to site $i$. From it, the average internal field is
$ 
h^{\mathrm{internal}} = \left\langle \frac{1}{N_s} \sum_i h^{\mathrm{internal}}_{i} \right\rangle.
$

In our $\pm J$ model, ${\cal F} = 1/2$ and $h^{\mathrm{internal}} = 0$  at saturation, because of the equal number of ferromagnetic and antiferromagnetic couplings.

The third rows of Figs.~\ref{fig:Figure_3_hardware_defined_spin_glass_hysteresis} and \ref{fig:Figure_4_2D_hysteresis_curves} show ``butterfly'' shaped hysteresis curves for the frustration ${\cal F}$. In both models, we see that ${\cal F}$ reaches a expected maximum of $0.5$ only where magnetization saturates to $1$, as predicted. At equilibrium, configurations at zero field minimize frustration, but this is generally not the case in  hysteretic, out-of-equilibrium regime. However,  both figures show that ar larger $\Gamma/J$, the system achieves considerably lower frustration, consistent with a more effective annealing from stronger quantum fluctuations.

Similarly, the fourth row of Figs.~\ref{fig:Figure_3_hardware_defined_spin_glass_hysteresis} and \ref{fig:Figure_4_2D_hysteresis_curves} plots the average internal field $h^{\mathrm{internal}}$ vs. $H_z$, which vanishes at full saturation, as predicted. Notably, the hysteresis structure of $h^{\mathrm{internal}}$ is more intricate than that of magnetization, displaying multiple pinch points. More generally, out-of-equilibrium anomalies such as non-monotonicities, asymmetries, switches, and memory are more pronounced in ${\cal F}$ and $h^{\mathrm{internal}}$ than in $M_z$.

In the full graph case (Fig.~\ref{fig:Figure_3_hardware_defined_spin_glass_hysteresis}), minimal frustration occurs near zero magnetization. In the hysteretic regime, this aligns with the coercive field, while in the quantum paramagnetic regime (large $\Gamma/J$), it instead coincides with zero field, where the butterfly collapses. In contrast, the square lattice case (Fig.~\ref{fig:Figure_4_2D_hysteresis_curves}) shows minimal frustration {\it before} coercivity, and interestingly, it aligns with the maximum absolute value of $h^{\mathrm{internal}}$.

Full saturation erases any memory of prior states, resulting in central symmetry (with respect to the origin) of the hysteresis curve. Instead, when the magnetization does not fully saturate, for minor loops this symmetry can be lost. This effect is evident in Fig.~\ref{fig:Figure_4_2D_hysteresis_curves},-(a), for $s=0.7$, and in Fig.~\ref{fig:Figure_3_hardware_defined_spin_glass_hysteresis},-(c), also for $s=0.7$, both of which exhibit a breakdown of central symmetry in their hysteresis loops. Interestingly, this asymmetry becomes more pronounced in the corresponding curves for ${\cal F}$.

By providing access to measured spin configurations, our methodology allows also extraction of magnetic structure factors (MSF)
$$
{S(q) = \langle \sum_{i, j} e^{i q \cdot (r_i - r_j)}  \sigma_z^i \sigma_z^j \rangle}
$$ 
($\mathbf{r}$ is measured in units of the lattice constant, the average is over 100 sampled spin configurations).
Fig.~\ref{fig:Figure_5_spin_structure_factor}a,b shows MSFs at different points of the hysteresis loops for the random bond model on a square lattice at $s=0.7$, across two machines~\footnote{For improved visualization clarity, Fig.~\ref{fig:Figure_5_spin_structure_factor} MSF plots used capped $\abs{S(q)}$ when plotting the heatmaps. Specifically, for all MSFs in panel (a), $\abs{S(q)}$ was capped at $8$, and for panel (b) they were capped at $\abs{S(q)}=6$.}. Panel (a) tells a familiar story of the fading of the central Bragg peak (at MSF 1), typical of the saturated state the central Bragg peak dominates (the surrounding weaker peaks on the axes, more apparent due to the logarithmic scale, are a finite size effect) as the system demagnetizes (MSFs 2–4), leading to the diffuse intensity of frustrated disorder, with residual short range ferromagnetic correlations.

Fig~\ref{fig:Figure_5_spin_structure_factor}(b) reports more distant points along the hysteresis loop, with increased structure in the demagnetized state. At MSF 1, intensity at the Brillouin zone boundary suggests emerging antiferromagnetic correlations atop weak ferromagnetic order, whereas, at MSF 2–4 fourfold symmetry breaking reflecting the disordered coupling, low edge intensity, and absent corner peaks suggest stripe-like rather than N{\'e}el antiferromagnetic order.

Fig.~\ref{fig:Figure_5_spin_structure_factor}(c) examines three points around a dip, labeled cyan, red, and green. While magnetization decreases from cyan to red, but then unexpectedly increases from red to green. $C(r)$, the correlation function along the $x$ and $y$ axes, whose undulated shape suggests a ``liquid'' of ferromagnetic domains, is effectively identical before and after the dip, suggesting reentrant ferromagnetic domains. This speculation is corroborated by real-space spin maps revealing larger ferromagnetic domains at these points. This reentrance is likely of quantum origin: the fraction of frustrated energy links ${\cal F}$, plotted in red on top of the magnetization curve, also shows a corresponding dip, signifying that  the ``internal energy", like the the longitudinal Zeeman energy, are reduced at the dip, but both increase as the longitudinal field is further reduced. The only possible energy gain to justify such a reentrant phase would be coming from the transverse Zeeman energy (the first term in Eq.~(\ref{equation:Transverse_Ising})). In the absence of projective measurements in the $x$ basis our hypothesis cannot currently be tested. Supplementary Information~\ref{section:appendix_non_monotonic_dip_SSF} examines this non-monotonic dip using averaged magnetic structure factors.

\section{Discussion}
\label{section:conclusion}

We have demonstrated for the first time that quantum annealers are viable platforms for the study of memory and hysteresis, and that their memory can be finely tune by controlling quantum fluctuations.
Our approach, particularized for D-Wave analog quantum computers, closely mimics laboratory protocols, but with the added, crucial advantage of a direct sampling of individual spin configurations, allowing for spatially resolved measurements of magnetization, internal fields, frustration, correlations, structure factors, and other essential measures for probing disordered and glassy systems, where bulk observables often obscure local dynamics and emergent behavior. We have showcased the rich phenomenology of our results to illustrate the broader potential of our method, though we leave to future works the full explanation of its many intricacies. 

Our methodology enables direct experimental access to Barkhausen noise and its potential fractal scaling in a tunable quantum system, as well as the investigation of training effects and limit cycles in hysteresis, phenomena central to Preisach-type models. In future works, we will explore precise probing of minor loops, return-point memory, ergodicity breaking, and spin glass aging, all characterized at the level of individual degrees of freedom, capabilities not accessible through previous approaches. Moreover, the sweep rate of the external field can be tuned, enabling controlled exploration of timescale effects in out-of-equilibrium dynamics (see Supplementary Information~\ref{section:appendix_protocol}). 

Beyond magnetism, the ability to engineer path-dependent responses has implications for neuromorphic computing, adaptive memory materials, artificial synapses, and associative memory models. More broadly, our results establish quantum annealers as experimental testbeds for non-equilibrium statistical mechanics, offering access to dynamical phenomena that are otherwise inaccessible to classical simulations or condensed matter experiments.

By reimagining quantum hardware as fully programmable laboratories for hysteresis and memory, this work sets a new benchmark for the scientific utility of quantum annealers and lays the groundwork for a new class of quantum computer enabled experiments.

\section*{Acknowledgments}
\label{sec:acknowledgments}
We thank Andrew King (D-Wave), Carleton Coffrin (LANL), Francesco Caravelli (Scuola Normale Superiore, Pisa), Vivien Zapf (LANL), Minseong Lee (LANL), Karin A. Dahmen (University of Illinois at Urbana Champaign) and Eugenio E. Vogel (Universidad de La Frontera) for discussions and helpful feedback on the manuscript.

{\bf Funding:} This work was performed under the auspices of the U.S. Department of Energy (DOE) at Los Alamos National Laboratory, operated by Triad National Security, LLC (contract 89233218CNA000001). It was supported by the Laboratory Directed Research and Development (LDRD) program at LANL under project numbers 20240032DR and 20240479CR-IST.
Computational resources were provided by the LANL Institutional Computing Program and the Darwin testbed, funded by the Computational Systems and Software Environments subprogram of LANL’s Advanced Simulation and Computing (ASC) Program (NNSA/DOE).
PS acknowledges support from the LANL Information Science and Technology Institute (ISTI) Fellowship. FB acknowledges support from the LANL Director’s Fellowship and the Center for Nonlinear Studies (CNLS) under program number 20250614CR-NLS.

{\bf Author Contributions:} EP devised the hysteresis protocol on quantum annealers, and performed all D-Wave experiments. EP and FB designed the experimental methods. EP, FB, PS, and CN developed the analysis. All authors contributed to data interpretation, and participated in experimental design. EP drafted the initial manuscript, and all authors contributed to the preparation and revision of the final version. CN supervised the project.

{\bf Competing Interests:} All authors declare no competing interests. 

{\bf Data and materials availability:} Data is available upon request to the authors.

%% file: appendix_sections.tex
\begin{figure*}[htb!]
    \includegraphics[width=0.49\linewidth]{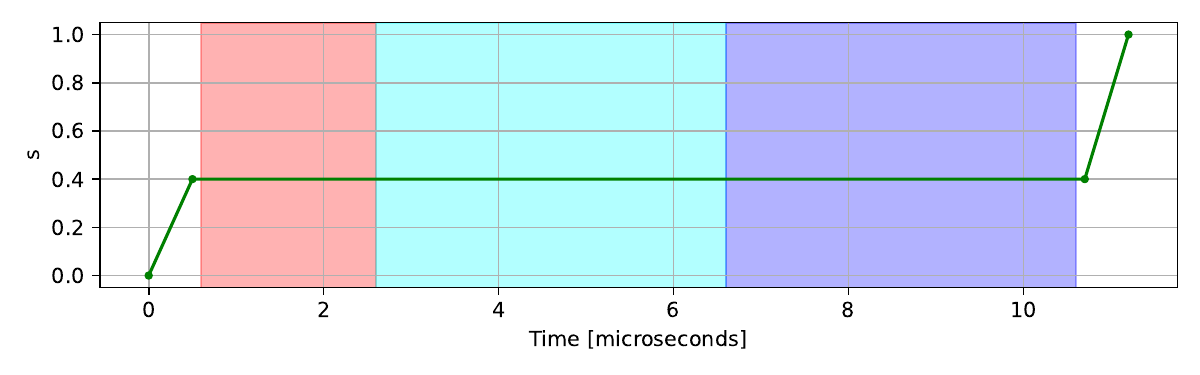}
    \includegraphics[width=0.49\linewidth]{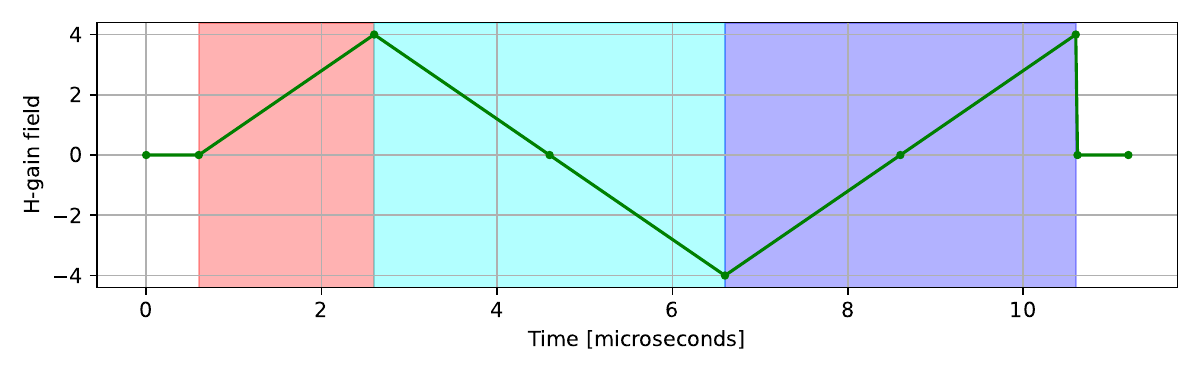}
    \caption{The time-dependent quantum annealing hardware schedules that define the hysteresis simulation protocol. The anneal schedule that defines the proportion of the transverse field relative to the classical diagonal Hamiltonian energy scale is given by values of $s$ specified at each point in time (left). The left-hand plot shows this anneal schedule for a specific example where the location of the anneal pause occurs at $s=0.4$. The right hand plot shows the h-gain (longitudinal) field time dependent protocol, which is a uniform multiplier on all local fields. This is a sampling based protocol where measurements are made at many slices of this continuous time waveform (specifically, at slices at progressive time intervals within the three color shaded regions), and then averaged statistics of these measurements can be extracted. These two plots are showing the complete schedules at the very end of the simulation where the h-gain sweep is complete - intermediate simulations use shorter annealing times along with quenches to $h=0$ before the qubit state measurements. Importantly, all non-zero h-gain fields are applied while the system is held at a specified transverse field proportion. The main regions of the simulation are marked as shaded vertical regions. Red shading shows the initial polarization ramp to get to the $+H$ polarized state; this is required to initialize the simulation since the hardware is initialized at $h=0$ and can have remnant magnetization or noise at $h=0$. Next, the cyan and blue shading show the two sweeps that implement one full (closed) hysteresis loop sweep, with the cyan going from $+H$ to $-H$ and the blue going from $-H$ to $+H$. We report the hysteresis data as the two sweeps that comprise the cyan and blue regions, neglecting the initial polarization ramp (red) as this is a initialization step. }
    \label{fig:hardware_schedules_appendix}
\end{figure*}

\begin{figure}[ht!]
    \includegraphics[width=0.999\linewidth]{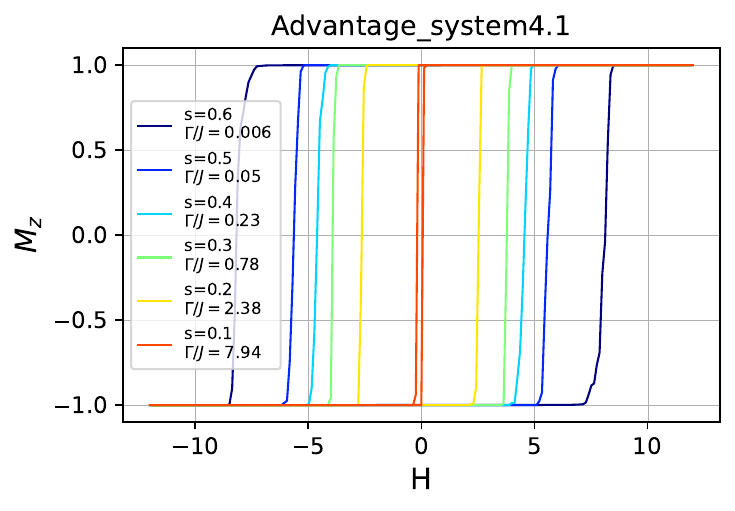}
    \caption{Magnetic hysteresis (Magnetization $M_z$, normalized to theoretical saturation, vs. Field $H_z$) of a hardware-graph ferromagnet run on \texttt{Advantage\_system4.1}. }
    \label{fig:hardware_3D_FM_4.1_appendix}
\end{figure}

\section{Magnetic Hysteresis Protocol on Programmable Quantum Annealers}
\label{section:appendix_protocol}

In this section, we describe the protocol in detail used for all our hysteresis experiments, implemented on various D-Wave QPUs. Table~\ref{table:hardware_summary} lists the three QPU's used in this study. 

Before going into the specific details of our protocol, we will first provide an overview of the annealing processes that are typically implemented on D-Wave's QPUs, including those that go beyond our specific protocols.
The time-dependent Hamiltonian implemented on D-Wave's QPUs is given by
\begin{align}
        {\mathcal H} =& - \frac{A(s)}{2}  \sum_i \hat\sigma_x^{i} 
        + \frac{B(s)} {2} \left( g(t) \sum_i h_i \hat\sigma_z^{i} + \sum_{\langle ij \rangle} J_{i j} \hat\sigma_z^{i} \hat\sigma_z^{j} \right).  
        \label{eq:H(s)_general}
\end{align}
Quantum annealing typically proceeds as follows.
All active qubits are initialized to the all up state in the Pauli-x basis. 
Next, the system evolves according to the dynamics dictated by the Hamiltonian in Eq.~\eqref{eq:H(s)_general}.
(For the annealing time scales used in this work, the dynamics is also influenced by environmental interactions.)
Finally, at the end of each annealing process, the state of each (active) qubit is measured in the computation basis. 
Typically, this process is repeated many time, so that the user gets access to many sampled spin configurations, one corresponding to each individual anneal process.

We note that $s$, which lies in the range $[0,1]$, is itself a user-specified function of time.
The functions $A(s)$ and $B(s)$, which have units of energy, denote the physical energy scale of the device.
While these two functions have qualitatively the same shape across various QPUs, their precise quantities differ from one QPU to another (see Supplementary Information~\ref{appendix:DWave_calibrated_schedules}). 

The user specified the coupler values $\{J_{i,j}\}$ and the local fields $\{h_i\}$ to encode a particular Ising model on the hardware. 
Typical annealing experiments correspond to $s$ ramping up linearly from $0$ to $1$, so that $s=t/t_\text{anneal}$, with $t$ denoting the annealing time.
While the standard quantum annealing protocol require $s$ to start and end at $0$ and $1$ respectively, it is possible to ``hold'' or ``pause'' $s$ at an intermediate value for most of the anneal.
We rely on this feature for all our experiments.
Another important user-defined control is the function $g(t)$, which is known as the $h$-gain schedule.
It controls the relative strength of the longitudinal field compared against the coupler strengths.
By default, it is typically set to $g(t)=1$ throughout the annealing process.
However, as described below, a time-varying $h$-gain schedule plays a key role in our hysteresis protocols.

The objective of the simulation is to emulate, using a probabilistic sampling-based quantum computational approach, a magnetic field sweep protocol that can reveal collective magnetic memory in a transverse-field Ising model. 
The simulation is primarily controlled by two elements. 
The first is the transverse field in the Ising Hamiltonian, which drives state transitions quantum mechanically.
(Note that due non-adiabaticity, along with thermal effects also cause transitions between different computational basis states.)
The second is (a time-varying) longitudinal field which can be chosen to have different values at different spins.
However, in all our simulations, we choose them to have the same value for all qubits.
For our hysteresis experiments, we vary this function (as a linearly interpolated schedule) in a periodic manner, with each repetition starting initially at $0$, then increasing it to the maximum positive value ($+H$), then to $-H$, followed by a return to $+H$. Here, the maximum applied $H$ field is dependent on the D-Wave QPU hardware properties. 

Importantly, the quantum annealers do not allow intermediate readout of qubit states in the middle of an anneal (when $\Gamma \neq 0$).
Consequently, continuous monitoring of the magnetization of the Ising model can not be performed - instead, we must incrementally prepare slices of this longitudinal field protocol and then measure the states of the qubits at various intermediate point in the protocol.
Additionally, because the sampling is probabilistic, for each schedule slice we perform multiple anneals and then average observable quantities are extracted for each parameter setting from the distribution of samples.
The average magnetization $M_z$, which is a measure of the overall alignment of spins in the lattice, is the primary quantity that we use in order to demonstrate the magnetic hysteresis protocol. When reporting $M_z$, we reverse its sign in order to compensate for the sign of the D-Wave Hamiltonian eq.~\eqref{eq:H(s)_general}. At the end of each anneal-readout cycle, the states of all of the (active) qubits are measured in the computational basis (i.e., the Pauli z basis denoted by $\sigma^z$).

Figure~\ref{fig:hardware_schedules_appendix} shows an example set of programmed schedules run on the D-Wave processor(s). These schedules are the two time-dependent control fields that facilitate the hysteresis simulation. This is effectively a two-sweep protocol (positive h-gain to negative h-gain, then negative h-gain to positive h-gain) that results in one closed hysteresis loop. During this changing h-gain field waveform, the system is paused at a fixed anneal fraction. The h-gain field is initialized at $0$; changing the h-gain field from $0$ to the maximum positive strength is required to initialize the two-sweep protocol, but we use this ramp to gradually polarize the state (as opposed to quenching this field rapidly to its maximum value). For this study, we use strictly the two-sweep technique shown in Figure~\ref{fig:hardware_schedules_appendix} -- but of course in principle one could use other sweep protocols. In practice we observe that the D-Wave quantum annealing hardware can frequently have remnant magnetization at initialization, which is why we use the initial longitudinal field ramp (shown by the red shaded region of Figure~\ref{fig:hardware_schedules_appendix}) to maximally polarize the system (up to the maximum longitudinal field we can apply on the hardware), before beginning the full sweep. Therefore, in practice we discard measurements during this initial ramp (red region), and only plot the closed hysteresis loop from the full sweep shown by the two blue regions. Some important specific aspects of the magnetic hysteresis simulation are defined, in detail, as follows:

\emph{Transverse field strength:} The parameter $s$ is the anneal schedule parameter that defines where in the standard anneal schedule the pause occurs (see~Eq.\eqref{eq:H(s)_general}), and is defined within the range $[0, 1]$. The term pause here refers to holding this hardware parameter $s$ constant as a function of the simulation time - this is illustrated by the left sub-plot of Figure~\ref{fig:hardware_schedules_appendix}. Smaller $s$ denotes stronger transverse field coupled with a weaker $J$, and larger $s$ denotes weaker transverse field coupled with stronger $J$. This parameter we vary typically over the range of $s=0.3$ to $s=0.7$ so as to observe differences when state transitions are easier or harder. At very smaller and very large $s$ values the hysteresis effect disappears. Importantly, this parameter $s$ couples together the fields $A(s)$ (the transverse field) and $B(s)$ (the programmed Ising model), meaning we can not, for example, independently vary $A(s)$ while keeping $B(s)$ fixed. This is a hardware restriction of the current D-Wave quantum annealers, but this means that varying $s$ is the only way to attenuate the transverse field strength. When reporting results, we typically report both this hardware defined normalized control parameter $s$ along with the ratio between the transverse field ($\Gamma$) and $J$. The energy scale of the transverse field present in the simulation is very important. Namely, if there is not sufficient transverse field ($s=1$), then there can be no state transitions and thereby no magnetization change. However, if there is only transverse field ($s=0$), then there is no magnetic memory to be probed because the system does not change.

\begin{table*}[ht!]
    \begin{center}
        \begin{tabular}{|l||l|l|l|p{2.6cm}|l|}
            \hline
            D-Wave QPU Chip & Graph name & Qubits & Couplers & Maximum h-gain Field Strength & Avg. Node degree \\
            \hline
            \hline
            \texttt{Advantage\_system4.1} & Pegasus $P_{16}$ & 5627 & 40279 & $\pm 3$ & $14.3$ \\
            \hline
            \texttt{Advantage\_system6.4} & Pegasus $P_{16}$ & 5612 & 40088 & $\pm 4$ & $14.2$ \\
            \hline
            \texttt{Advantage2\_prototype2.6} & Zephyr $Z_{6, 4}$ & 1248 & 10827 & $\pm 1.75$ & $17.6$ \\
            \hline
        \end{tabular}
    \end{center}
    \caption{Summary of the D-Wave quantum processing units (QPUs) used in this study. The average node degree refers to the average connectivity of the (undirected) hardware graph, where nodes represent qubits and edges represent couplers. 
\label{table:hardware_summary}}
\end{table*}

\emph{Programmable Local Fields:} Whether the uniform local fields, for all active qubits, are set to all positive or all negative coefficients. The sign of these local fields should not significantly change the protocol - but could result in slight differences due to hardware noise or control errors. These local fields are what the h-gain time dependent control field acts on (as a multiplicative factor). In practice, in order apply the overall strongest longitudinal field possible (complemented by the time dependent multiplier of the h-gain field), we set this local field strength to the largest value that is allowed to be programmed on the hardware. Note that other patterns of local fields could be programmed on the D-Wave hardware, but generally we assume a uniform field will be applied with the goal of emulating the physical protocols that would be performed in a experimental hysteresis sweep on a real material. For all simulations we present in this study, we apply a uniform positive coefficient field. The use of the local field programmability (setting the $h_i$'s), alongside the h-gain field, means that the Ising model that we probe with the hysteresis protocol must be defined entirely by the $J_{i, j}$ coupling terms.

\emph{Longitudinal Field Sweep Strength:} This parameter is the strongest h-gain field (longitudinal field) that is applied during the sweep protocol\footnote{This field is referred to as \emph{h-gain} in the D-Wave hardware control parameters, but we will refer to this interchangeably as the \emph{longitudinal field} or the \emph{h-gain field}}. This field is specified by the function $g(t)$ in Eq.\eqref{eq:H(s)_general}. The applied field strength for both the positive and negative sign is always symmetric during the sweep (as illustrated by Figure~\ref{fig:hardware_schedules_appendix}-right). In principle, longitudinal fields that are weaker than the maximum allowable field that can be programmed on the hardware could be used, for example when studying minor loops. In practice, we use the strongest h-gain field that can be programmed on each D-Wave QPU. The (maximum) strength of the applied h-gain field is very important for this protocol, namely because ideally the simulation would reach full magnetic saturation -- however if the longitudinal field is not strong enough relative to the magnetic system encoded on the hardware, then full saturation can not be reached (Table~\ref{table:hardware_summary} lists the maximum longitudinal field that can be applied for each D-Wave device).

\emph{Longitudinal Field Points}: As illustrated in Figure~\ref{fig:hardware_schedules_appendix}-right, the way that the h-gain schedule is defined is by a series of points which are then linearly interpolated between. At a minimum, these schedules used in this study require $7$ points, but more complex schedules would need more h-gain schedule points (note that this is also true for more complex annealing schedules). The D-Wave QPU's used in this study allow only a maximum of $20$ h-gain schedule points (and a maximum of $12$ anneal schedule points) to be programmed by the user.

\emph{Anneal Schedule Ramps:} How long the anneal schedule ramps take to reach the $s$ anneal fraction, with the system having been initialized at $s=0$ with maximum transverse field, and then similarly to go back to the $s=1$ anneal fraction for qubit readout. The D-Wave hardware requires that the qubits are readout at $s=1$ where there is no transverse field being applied. Ideally this change to $s=1$ would be done nearly instantaneously, but in practice there is a constraint on how fast the anneal schedule can be changed meaning that these ramps do require up to $0.5$ microseconds of simulation time. Although a required part of the simulation, we neglect these quenches in the hysteresis protocol in the sense that we do not measure at intermediate times during these quenches, and moreover we ensure that the total simulation times are larger than these quenches ensuring that a majority of the dynamics we observe are due to the pulsed longitudinal field. In practice, we use a ramp duration of $500$ nanoseconds for all simulations both for the initial quench ramp and for the final readout ramp. 

\emph{Longitudinal Field Ramp Decrease:} How fast the h-gain field is reduced to $0$ field during the intermediate slices of the protocol. Ideally this change would also be done nearly instantaneously. In practice, we set this duration to be $20$ nanoseconds for all simulations to adhere to the maximum slope requirements on the D-Wave hardware for this control field. We want to set this field to zero during readout, and in particular during the hardware-required ramp immediately preceding readout, because otherwise the longitudinal field would continue to be applied during this ramp (e.g., a changing transverse field), instead of a static transverse field.

\emph{Turning Off the Longitudinal Field Pre-readout:} This parameter defines how long we allow for the system to equilibrate after the initial anneal schedule ramp to the target pause at a given $s$ value, and after the last step of the h-gain field is applied. We set this duration to be quite fast at $0.1$ microseconds so as to minimize the effect of the transverse field while the longitudinal field is not applied. It is during this $0.1$ microsecond anneal-schedule pause that the longitudinal field is turned off via a rapid quench of $20$ nanoseconds (described above). Longer equilibration times could be applied to this protocol, but this would result in the transverse field acting more on the system after the longitudinal field sweeps -- in particular this causes more relaxation of the system towards a demagnetized state We note that simulations with longer equilibration times do exhibit magnetic hysteresis (at least with up to tens of microseconds of pause time), but for the simulations shown in this study we focus on very short equilibration times with the goal of attempting to readout the state of the qubits close to the removal of the longitudinal field. For symmetry of the schedules, we also use this same duration of $0.1$ microseconds of additional pause time, after the initial $0.5$ microsecond quench to the target s, to wait before turning on the longitudinal field.

\begin{figure*}[th!]
    \centering
    \includegraphics[width=0.88\linewidth]{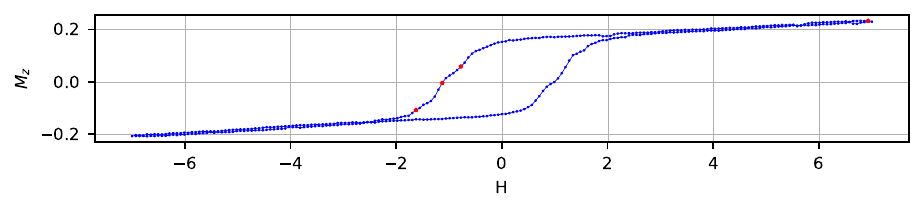}\\
    \includegraphics[width=0.246\linewidth]{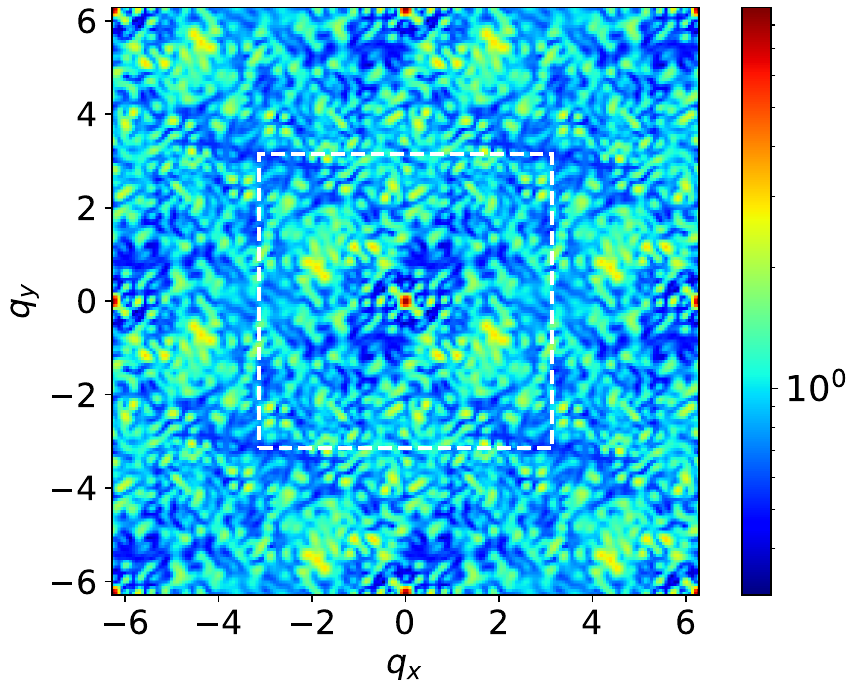}
    \includegraphics[width=0.246\linewidth]{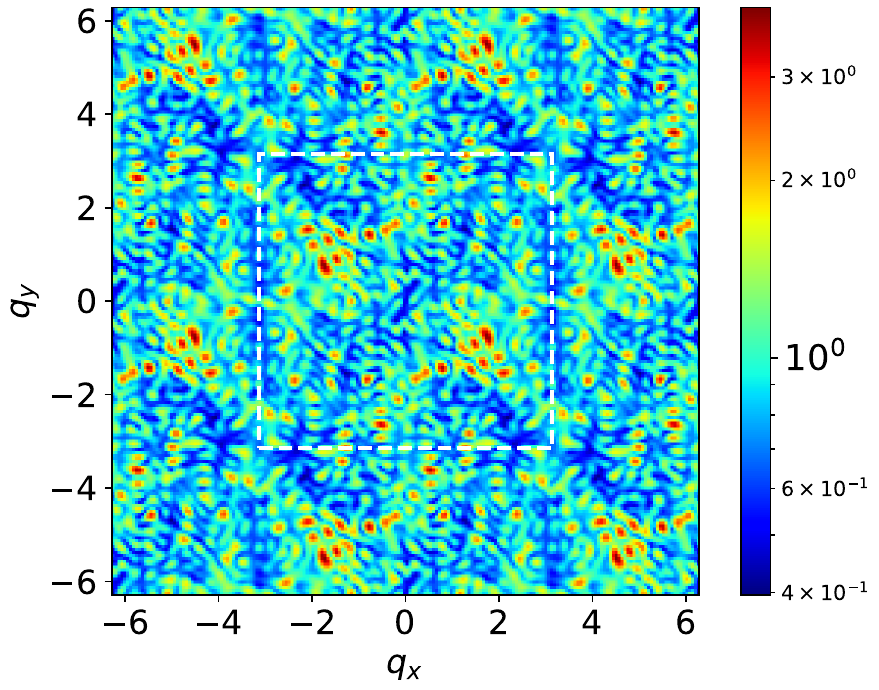}
    \includegraphics[width=0.246\linewidth]{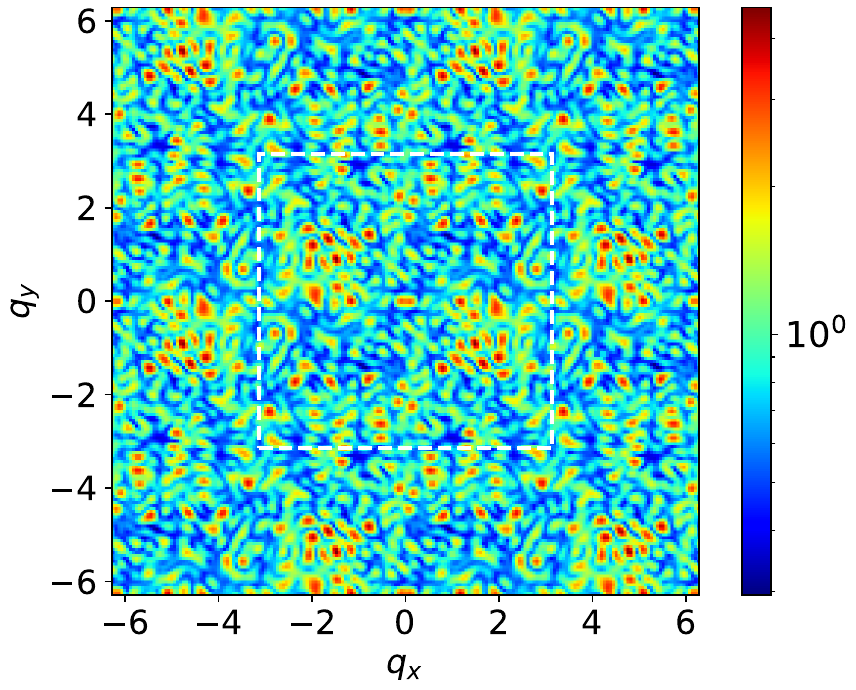}
    \includegraphics[width=0.246\linewidth]{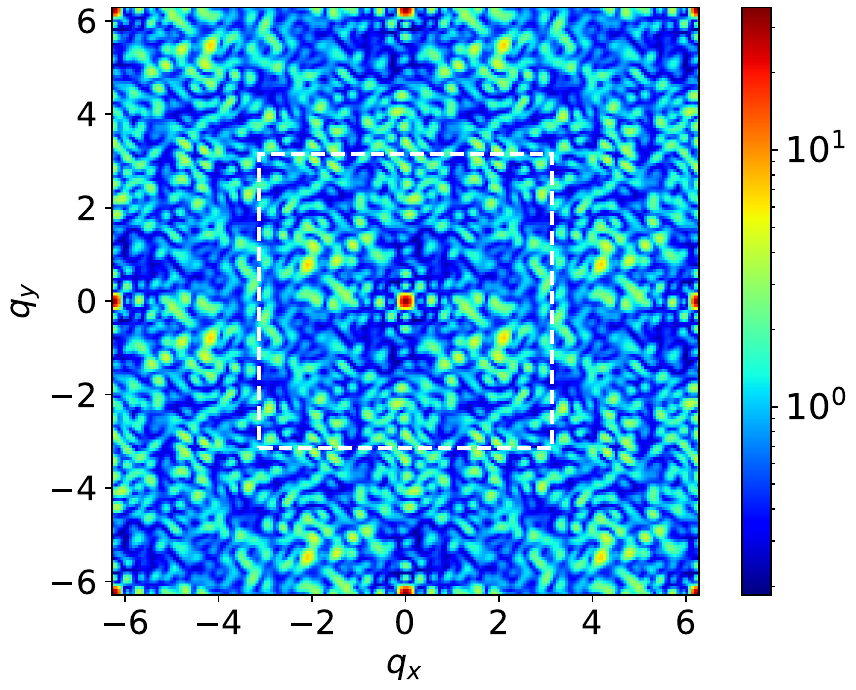}
    \caption{Magnetic spin structure factor $\abs{S(q)}$ heatmaps (bottom), averaged over the first $100$ spin configurations sampled on the QPU, at representative points along the hysteresis cycle (top) for a 2D square grid ($26\times 26$ spins) $\pm J$ model run on the \texttt{Advantage2\_prototype2.6} processor at $s=0.3$. The order of the red points on the hysteresis loops correspond to the order of the MSF plots shown below the loop. Each MSF heatmap uses a scale determined by that $\abs{S(q)}$ matrix, in other words the heatmap scale is not the same across the four heatmaps. The dashed white box outlines the first Brillouin zone. }
    \label{fig:SSF_2D_averaged_zephyr_s0.3}
\end{figure*}

\begin{figure*}[th!]
    \centering
    \includegraphics[width=0.88\linewidth]{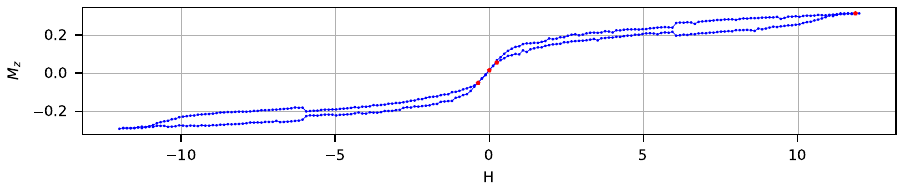}\\
    \includegraphics[width=0.246\linewidth]{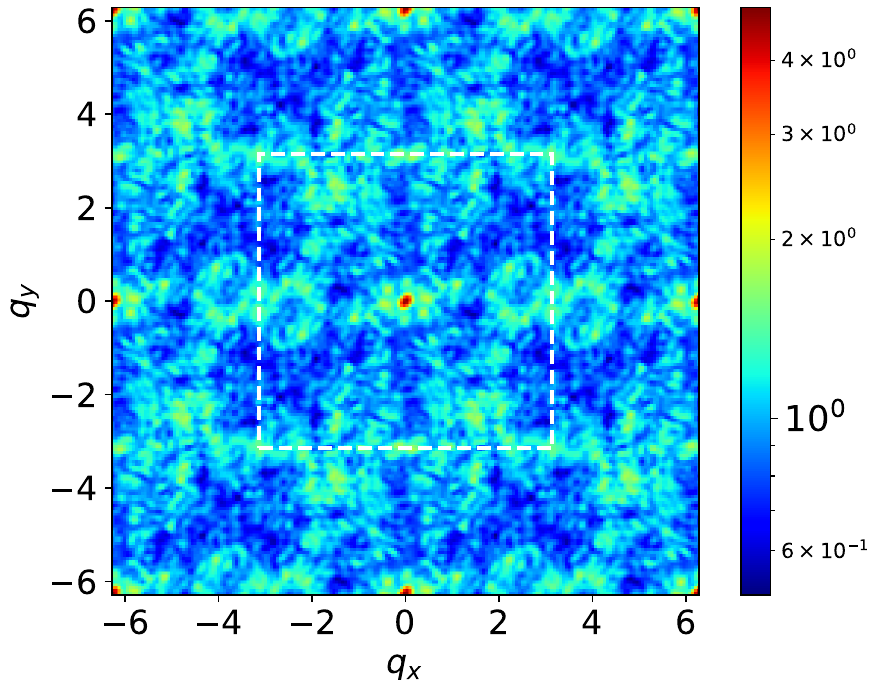}
    \includegraphics[width=0.246\linewidth]{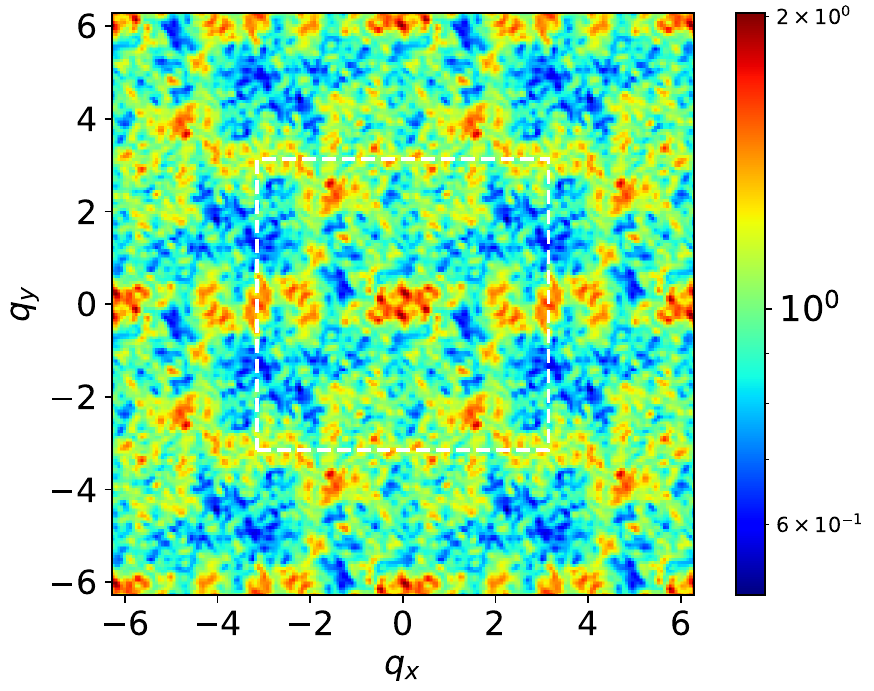}
    \includegraphics[width=0.246\linewidth]{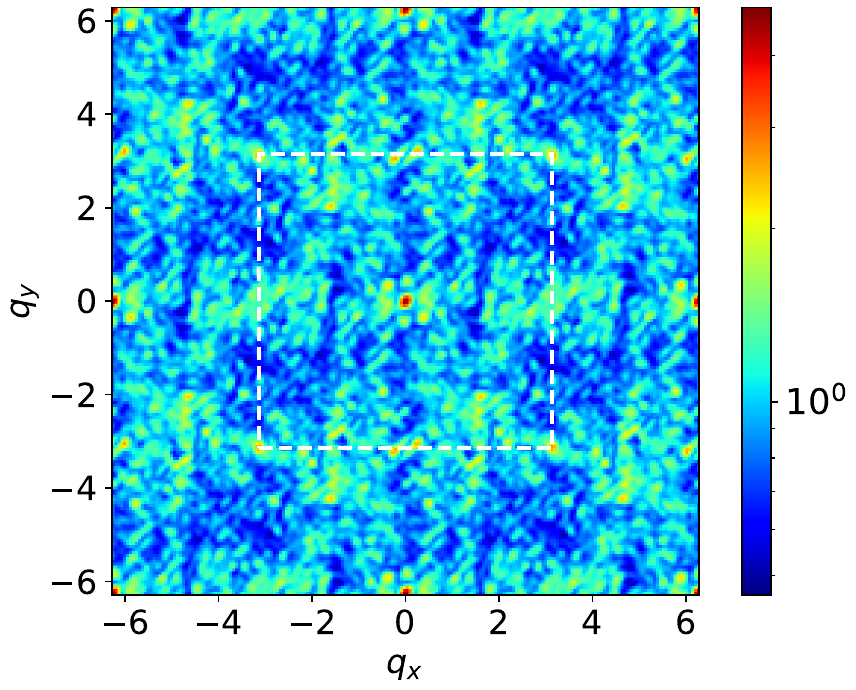}
    \includegraphics[width=0.246\linewidth]{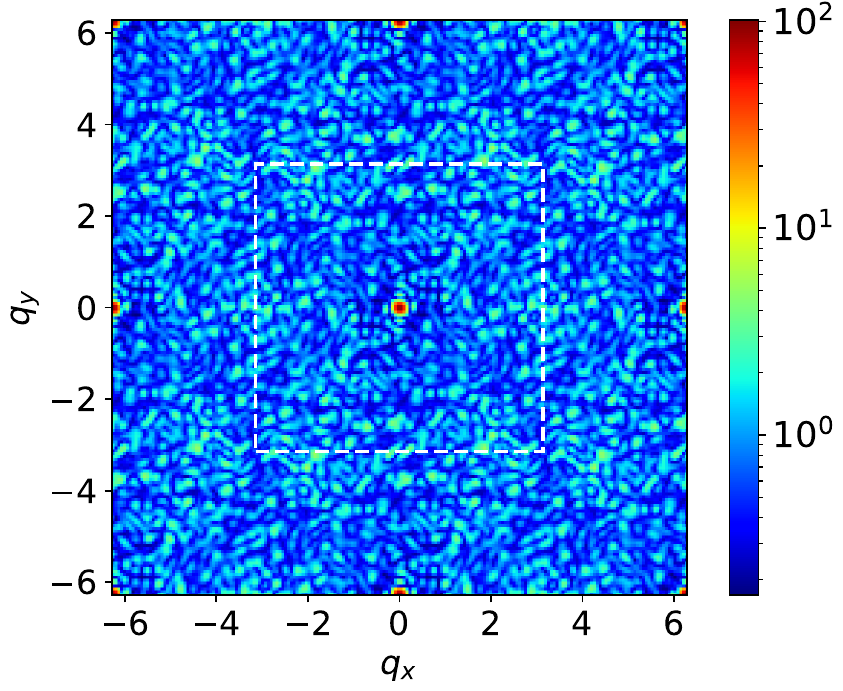}
    \caption{Magnetic spin structure factor $\abs{S(q)}$ heatmaps (bottom), averaged over the first $100$ spin configurations sampled on the QPU, at representative points along the hysteresis cycle (top) for a 2D square grid ($32\times 32$ spins) $\pm J$ model run on the \texttt{Advantage\_system4.1} processor at $s=0.3$. The order of the red points on the hysteresis loops correspond to the order of the MSF plots shown below the loop. Each MSF heatmap uses a scale determined by that $\abs{S(q)}$ matrix, in other words the heatmap scale is not the same across the four heatmaps. The dashed white box outlines the first Brillouin zone.  }
    \label{fig:SSF_2D_averaged_pegasus_s0.3}
\end{figure*}

\emph{Total Simulation Time:} The total length of the simulation can be changed quite significantly -- the only requirement is that there is sufficient time to allow the h-gain sweep protocol to be implemented on the hardware and several points of that sweep to be sampled from. In D-Wave device parameter terminology this parameter is known as the annealing time, however in this case the process is not doing annealing, so we often interchangeably refer to this parameter as the simulation time. The minimum programmable annealing time on these D-Wave QPUs is $500$ nanoseconds and the longest allowed annealing time is $2000$ microseconds \footnote{These QPUs do actually allow faster annealing times, down to 5 nanoseconds, but in that regime modified transverse field schedule control as well as longitudinal field control is not supported and therefore this protocol can not be applied in the anneal time regime of less than $500$ nanoseconds. }. The short equilibration time and ramp durations combined have a simulation time of $1.2$ microseconds. For all experiments reported in this study we use an annealing time of $11.2$ microseconds. Note that this is the total annealing time at the very end of the simulation, intermediate steps of the simulation have a shorter total annealing time - for example the very first ``slice'' of the h-gain field sweep (post-maximum polarization) would use a total simulation time of $\sim 3.2$ microseconds. 

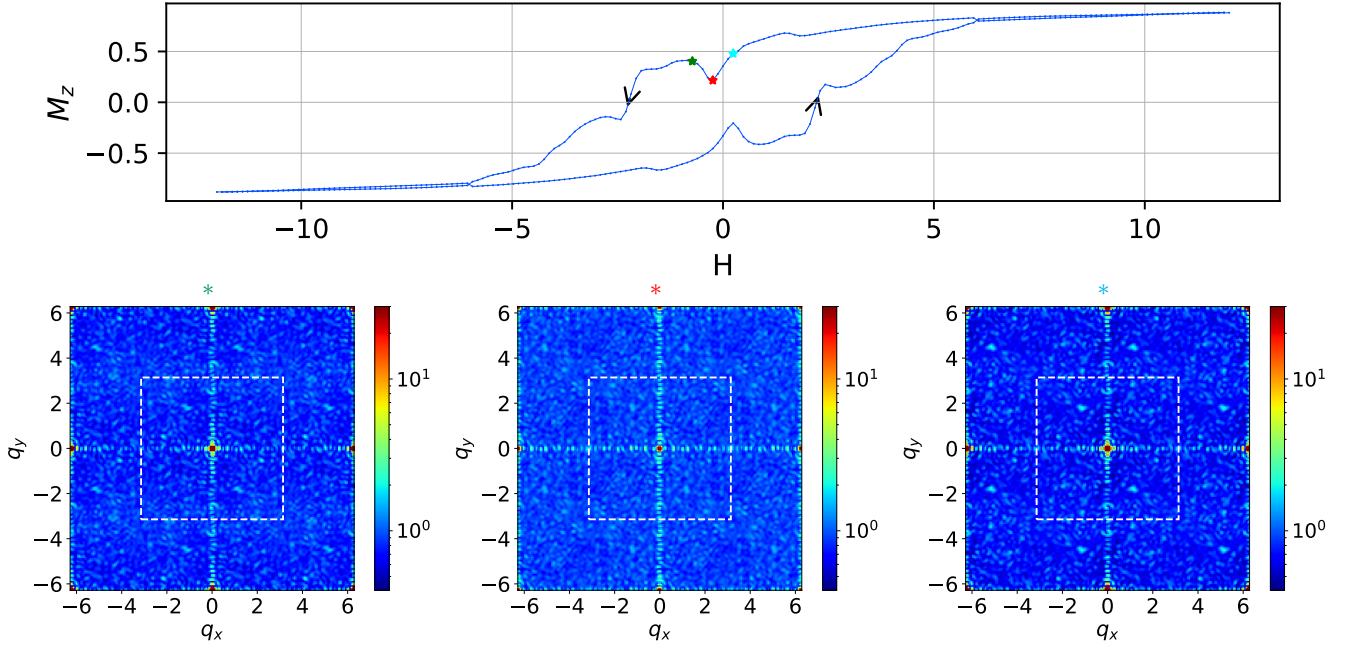
\begin{figure*}[ht!]
    \centering
    \includegraphics[width=0.94\linewidth]{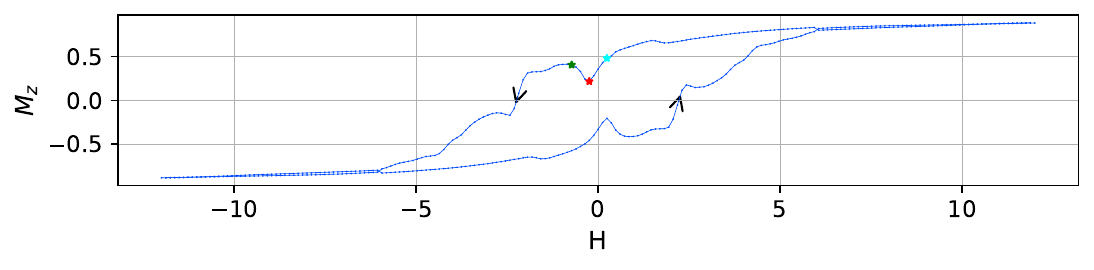}\\
    \begin{picture}(0,0)
        \put(78,130){\textcolor{ForestGreen}{*}}
    \end{picture}
    \includegraphics[width=0.32\linewidth]{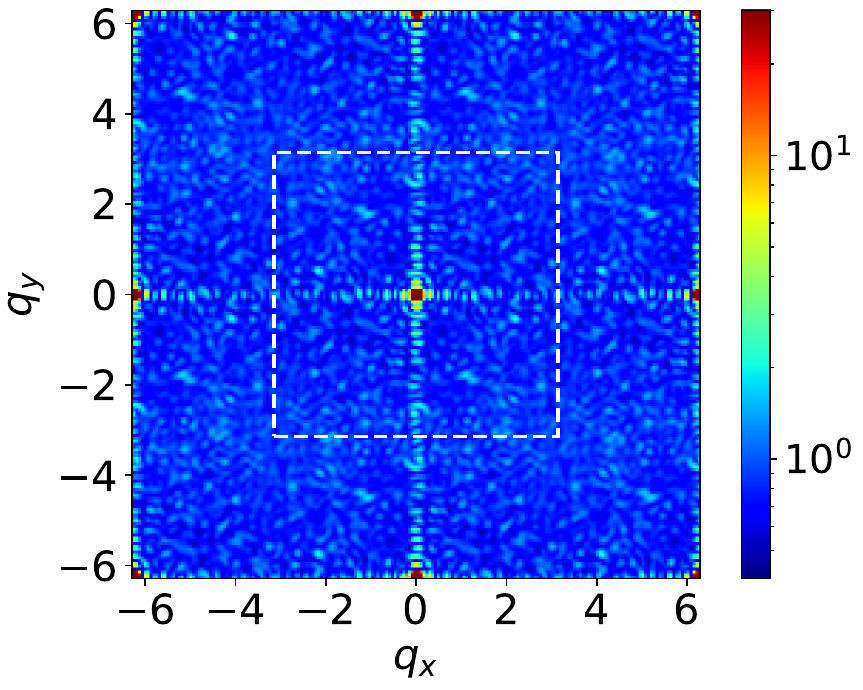}
    \begin{picture}(0,0)
        \put(78,130){\textcolor{red}{*}}
    \end{picture}
    \includegraphics[width=0.32\linewidth]{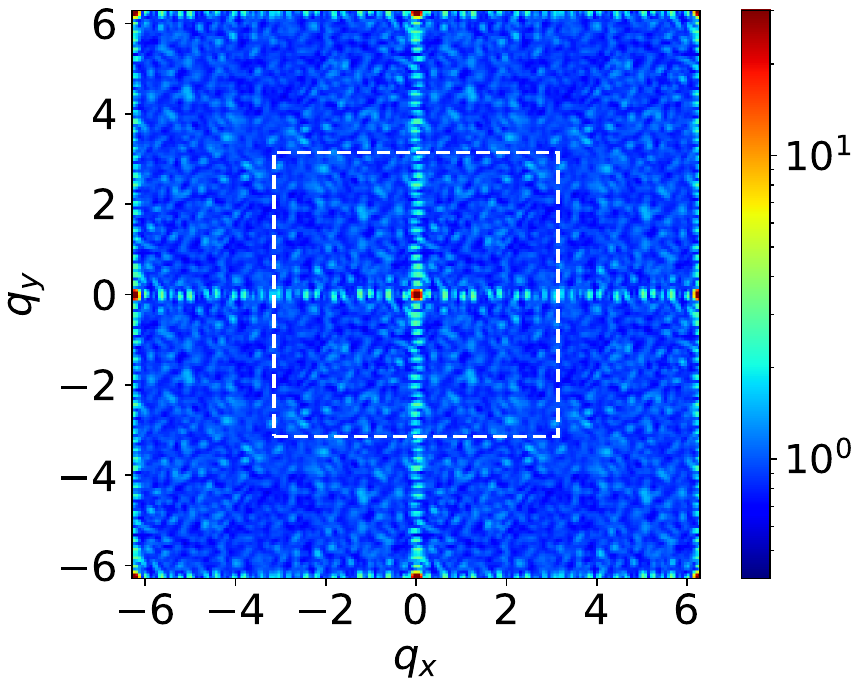}
    \begin{picture}(0,0)
        \put(78,130){\textcolor{cyan}{*}}
    \end{picture}
    \includegraphics[width=0.32\linewidth]{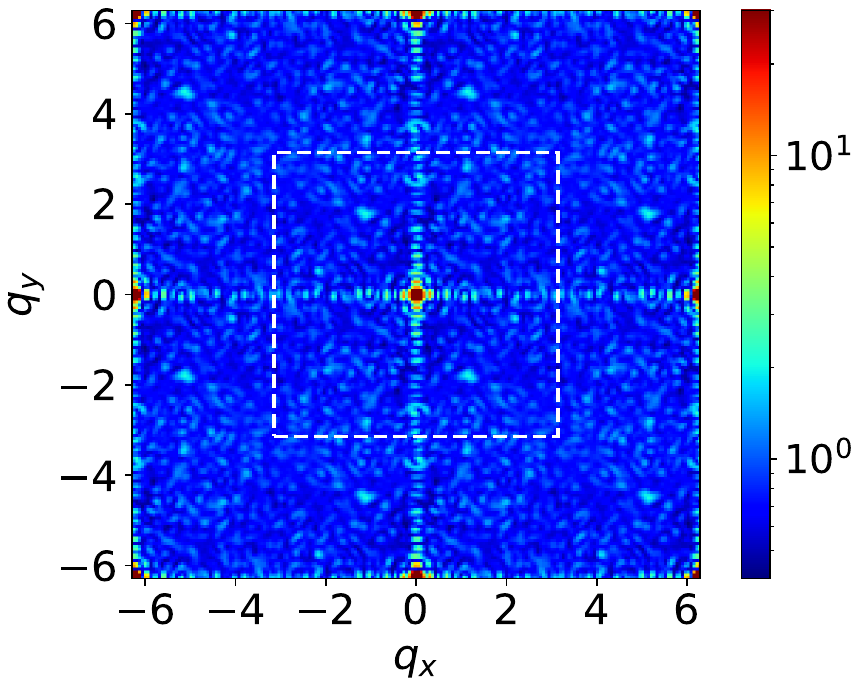}    
    \caption{This figure reports hysteresis data from a $32\times 32$ $\pm J$ model run on \texttt{Advantage\_system4.1} at $s=0.6$ along with three MSFs (cyan point is before the dip, red point is at the dip, and green point is after the dip). The goal here is to probe the structure factor around and at a very clear non-monotonic magnetization dip. All MSF plots use a $\abs{S(q)}$ heatmap on a log scale. The dashed white box outlines the first Brillouin zone. The three points are color-coded as green, red, cyan. }
    \label{fig:appendix_focused_example_non_monotonic_dip_SSF_plots}
\end{figure*}

\begin{figure*}[th!]
    \centering
    \includegraphics[width=0.246\linewidth]{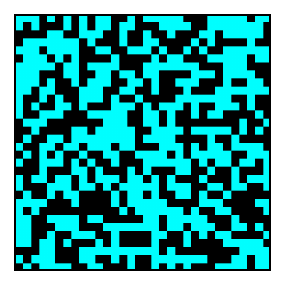}
    \includegraphics[width=0.246\linewidth]{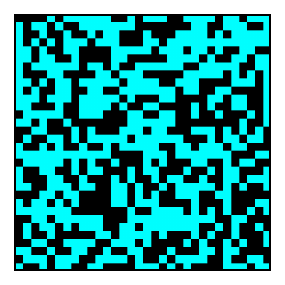}
    \includegraphics[width=0.246\linewidth]{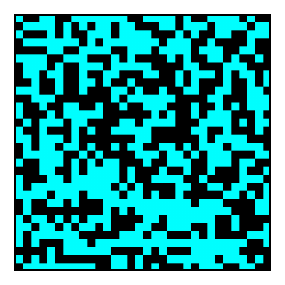}
    \includegraphics[width=0.246\linewidth]{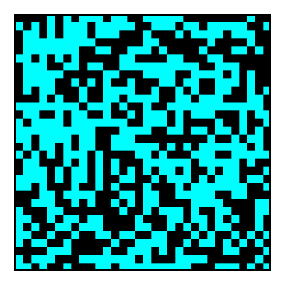}
    \caption{Four representative example plots of individual samples (spin configurations) from the $32\times 32$ $\pm J$ grid Ising model, generated on the \texttt{Advantage\_system4.1} processor at $s=0.7$, in the approximately de-magnetized region of the forward longitudinal sweep. In particular, these spin configurations come from an applied longitudinal field of $-5$ (in D-Wave hardware normalized units), with an average net magnetization $M_z$ of $\approx 0.05$ across these four individual samples. Cyan pixels denotes a spin up $+1$ and black pixels denotes a spin down $-1$. }
    \label{fig:real_config_2D_plots_4.1_s0.7}
\end{figure*}

\begin{figure*}[th!]
    \centering
    \includegraphics[width=0.49\linewidth]{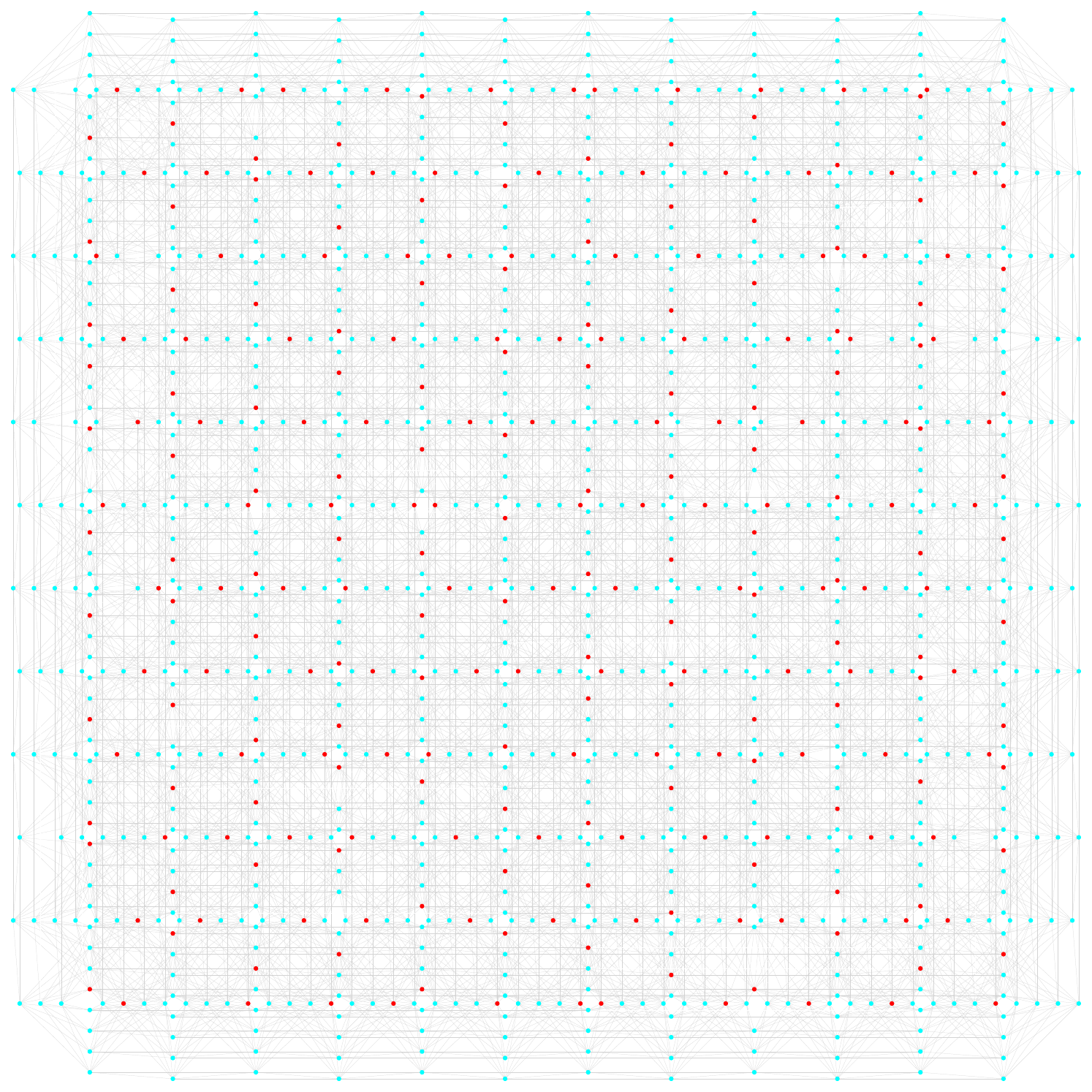}
    \includegraphics[width=0.49\linewidth]{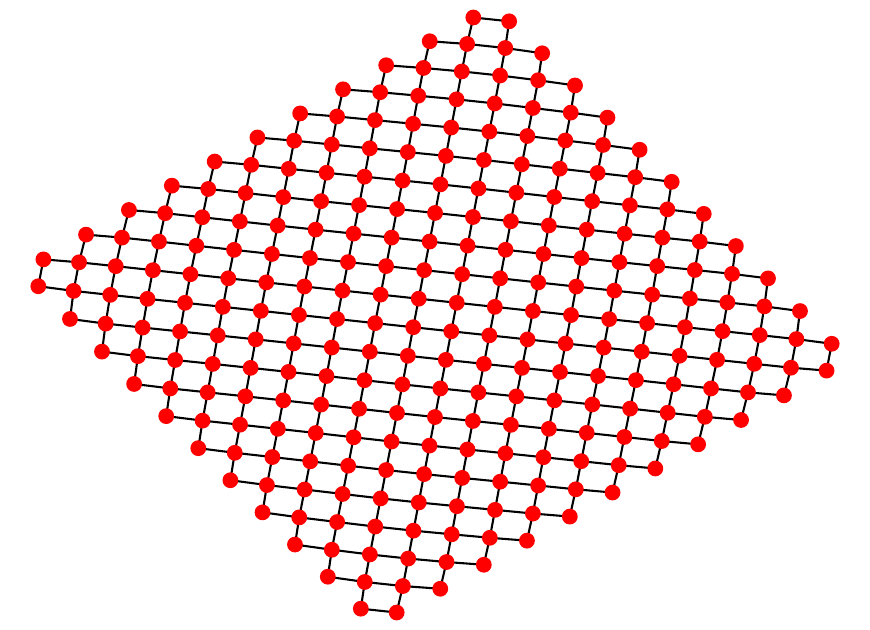}
    \caption{Planar subgraph of the \texttt{Advantage2\_prototype2.6} Zephyr hardware graph. Left: the full hardware graph of the D-Wave QPU, where nodes are qubits and edges connecting them are hardware couplers. All couplers are shown as grey lines. Red denotes qubits that form the strictly planar 2D square grid subgraph, and cyan denotes all other qubits. Note that the hardware has couplers and qubits that are missing from the logical $Z_{6, 4}$ graph. Right: the induced subgraph defined by the red qubits in the left hand hardware rendering. This subgraph is strictly a planar subgraph -- there exist no other connections between these qubits besides those shown in the network rendering on the right. There are also other possible planar subgraphs of the hardware graph - here we use this single square grid embedding for all spin structure factor plots of the full hardware-defined spin glass models. The planar subgraph (right) contains exactly 264 nodes, whereas the hardware graph contains exactly 1248 nodes. }
    \label{fig:zephyr_graph_and_planar_subgraph}
\end{figure*}

\emph{Number of Sampled Points Along the Longitudinal Field Sweep:} Because this protocol is a \emph{sampling} based protocol, we need to select some number of points along the longitudinal field sweep to measure. For all simulations reported in this study, an annealing time of $11.2$ microseconds is used, with $\approx 500$ linearly spaced points along the longitudinal field sweep (this is $100$ points along the five linear segments of the longitudinal field ramps in Figure~\ref{fig:hardware_schedules_appendix}, and a total of $400$ points that form the closed hysteresis loops), and a total of $2000$ (independent) anneal-readout cycles for each point along the hysteresis loop. $2,000$ samples for each point gives negligible finite sampling effects (e.g., shot noise) on the observables that we quantify such as $M_z$, but it is of course still present in these simulations -- reducing shot noise more significantly could be an important consideration for simulations that are especially sensitive to the precision of measured observables. This number of linearly spaced points of $\approx 500$ is motivated by the minimum annealing time resolution of $0.01$ microseconds on D-Wave QPU's -- for any other total simulation times that one could use in this protocol, the total number of sampled points should not cause the annealing time difference to drop below the minimum annealing time resolution of the hardware.

All other programmable D-Wave hardware parameters not specified here are left at default values for the simulations that we report. Future studies could make use of statistic balancing calibrations of hardware parameters such as flux bias offsets so as to fine tune the simulations.

\section{Ferromagnetic Model Hysteresis on \texttt{Advantage\_system4.1}}
\label{section:appendix_remaining_ferromagnetic_model_hysteresis}

Figure~\ref{fig:hardware_3D_FM_4.1_appendix} shows a complete set of hysteresis cycles on a hardware-graph defined ferromagnet, run on \texttt{Advantage\_system4.1}. Compared with Figure~\ref{fig:Figure_2_hardware_3D_FM_area}, these hysteresis loops look nearly identical despite slight hardware differences. 

The simulation of the fully ferromagnetic model defined on every coupler of the hardware graph of \texttt{Advantage2\_prototype2.6} we do not show because the maximum possible longitudinal field strength that can be programmed on \texttt{Advantage2\_prototype2.6} is weaker than the other two devices, which (combined with a slightly higher average qubit degree, see Table~\ref{table:hardware_summary}) resulted in the field being too weak to induce a magnetization reversal (meaning, the net magnetization, post initial polarization ramp, was constant).

\begin{figure*}[th!]
     \centering
     \includegraphics[width=0.97\linewidth]{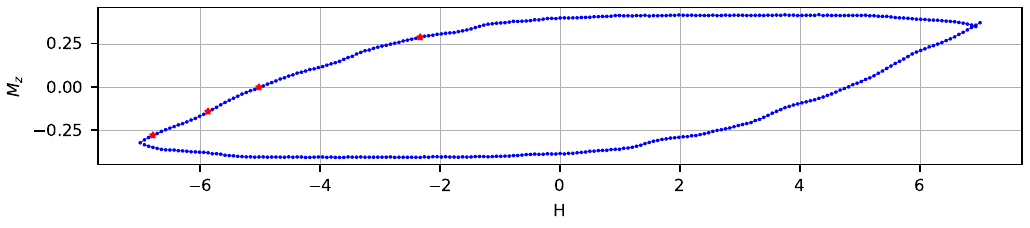}\\
     \includegraphics[width=0.246\linewidth]{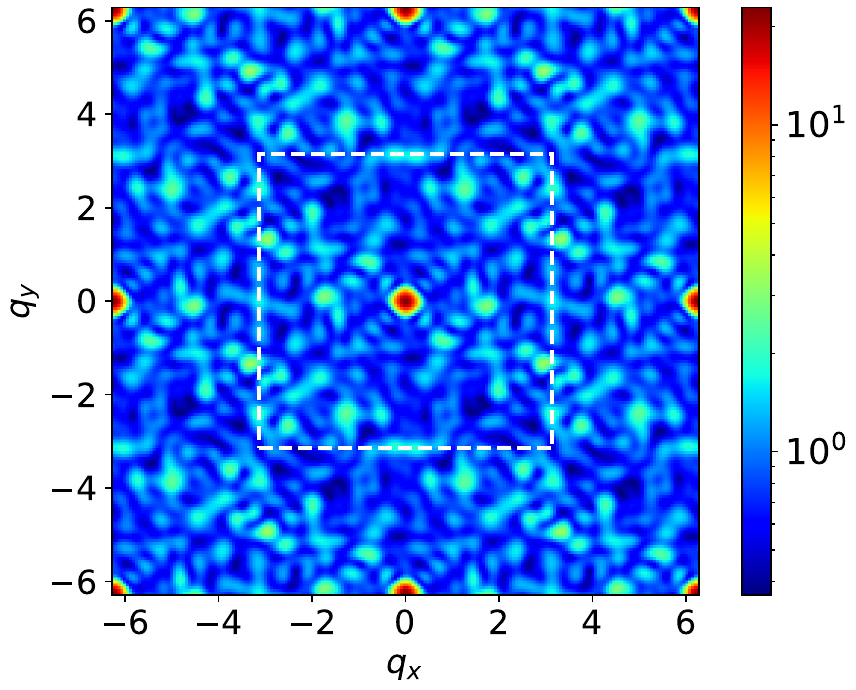}
     \includegraphics[width=0.246\linewidth]{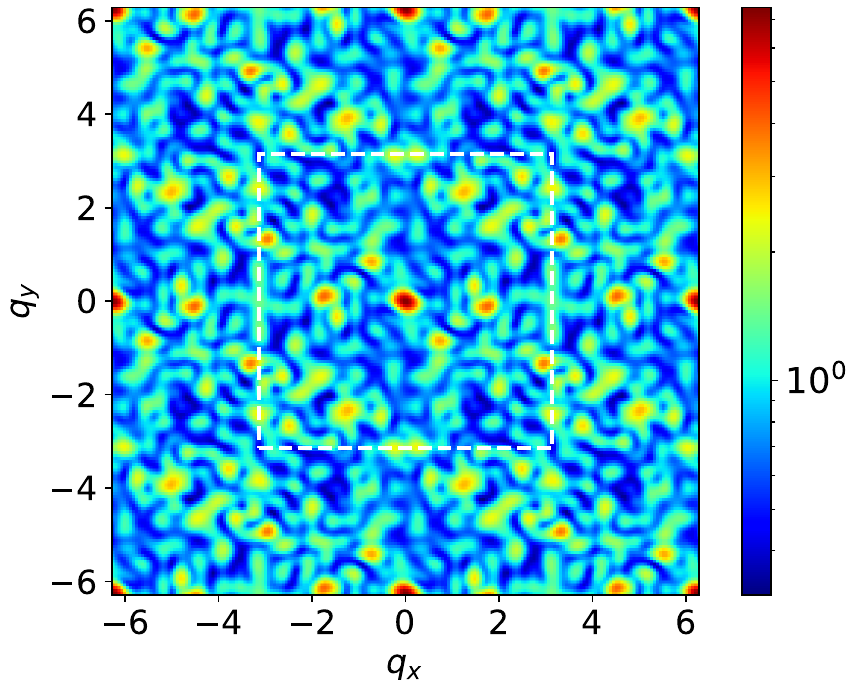}
     \includegraphics[width=0.246\linewidth]{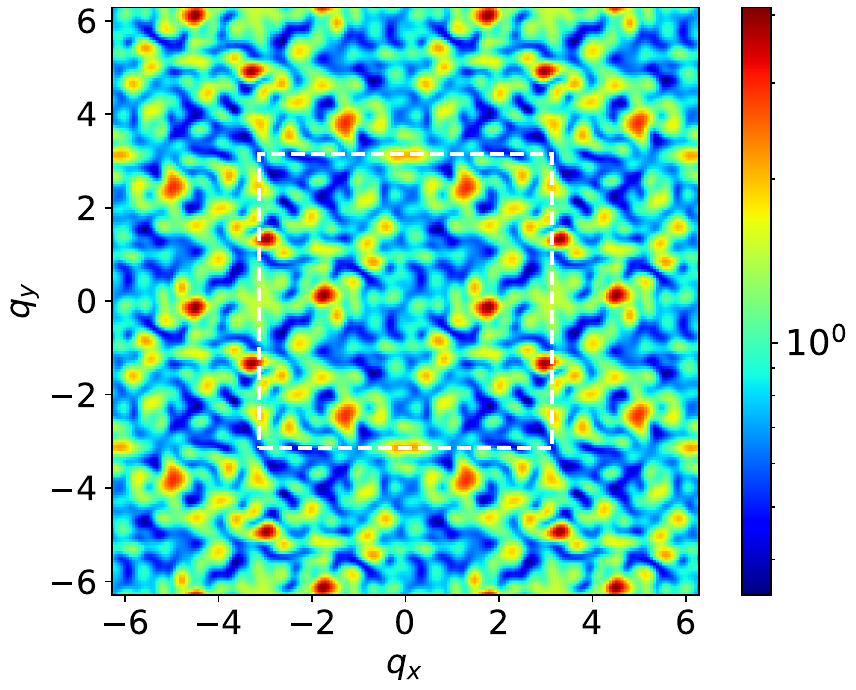}
     \includegraphics[width=0.246\linewidth]{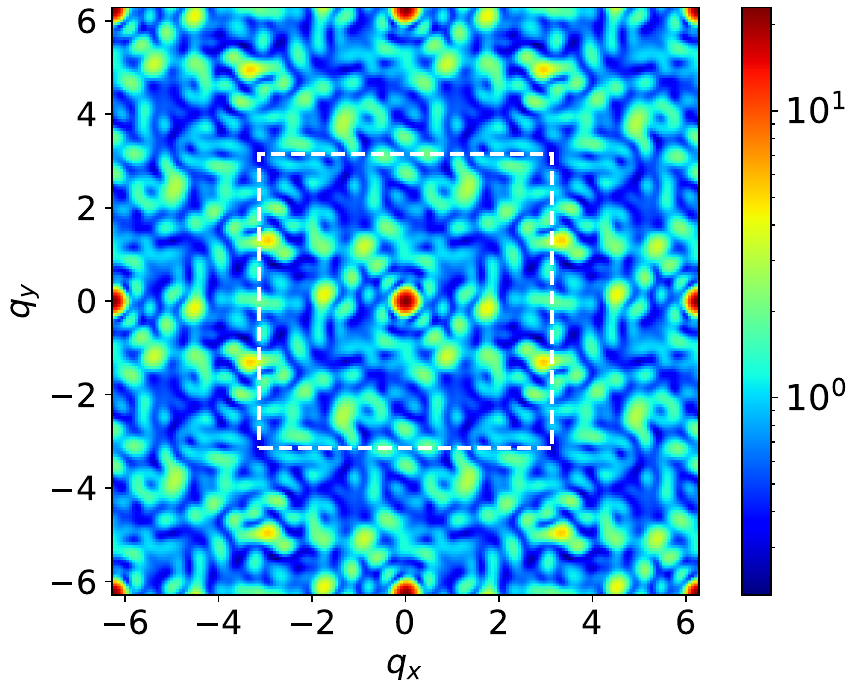}
     \caption{Averaged magnetic spin structure factor $\abs{S(q)}$, in log scale, from the quantum annealing hysteresis protocol at $s=0.7$ (weaker transverse field, and stronger $J$) from a $264$ node planar grid subgraph (e.g., 2D slice) of the Zephyr hardware graph \texttt{Advantage2\_prototype2.6} defined $\pm J$ model at four specific points along the hysteresis cycle. The hysteresis curve on the top plot is the average single site magnetization $M_z$ across the entire lattice, and the red asterisks denote specific points during the hysteresis cycle which we have extracted averaged MSFs of. The averaged MSF at each one of those points is given at the bottom - the order of these plots corresponds to the order of the red points on the hysteresis curve. Each MSF heatmap is averaged over the first $100$ spin configurations measured on the quantum annealing hardware for progressively longer-duration longitudinal field slices. Here, the four MSF plots do not share the same heatmap scale. The dashed white box outlines the first Brillouin zone. }
     \label{fig:SSF_Zephyr_Slice_of_3D_spin_glass_averaged_s0.7}
\end{figure*}

\section{Averaged 2D Ising Model Magnetic Spin Structure Factor Heatmaps}
\label{section:appendix_averaged_magnetic_spin_structure_factor}

We plot the magnetic spin structure factor (MSF) for momentum vector $\vec q$ computed at locations in a $200\times 200$ uniformly spaced grid spanning $(-2\pi,2\pi)$. 
The magnetic spin structure factor $S(\vec{q})$ can be computed for each spin configuration separately.
When computing $S(\vec{q})$, we set that the lattice spacing is $1$.
For all our MSF plots, we compute and plot $\abs{\langle S(q) \rangle}$ as a function of $\vec{q}$, where $\langle \circ \rangle$ denotes an averaging over $100$ samples. This is motivated by averaging over a non-negligible number of samples to find if there are consistent correlations between spins during the hysteresis simulation. In particular, having too few samples causes the spin structure factor to suffer from finite sampling effects where there can be concentrations of the $S(q)$ vector by chance rather than showing a true underlying structure. We find $100$ configurations is sufficient to average out finite sampling and finite system size effects. Magnetic structure factor allows us to visually examine the types of magnetic ordering that occur during these hysteresis cycles.

Figure~\ref{fig:SSF_2D_averaged_zephyr_s0.3} displays a hysteresis loop and selected magnetic spin structure factor plots, corresponding to specific points along the hysteresis cycle, for a $\pm J$ model on a $26\times 26$ 2D square lattice, implemented on \texttt{Advantage2\_prototype2.6} for an anneal fraction of $s=0.3$. 
As the maximum longitudinal field strength is comparatively weak and the transverse field is strong at $s=0.3$, we don't see magnetic saturation. 
Examining the MSF corresponding to the demagnetized regime (with MSF organized left to right corresponding to the points on the hysteresis loop left to right) we see weak peaks away from the center of the Brillouin zone indicating antiferromagnetic stripe ordering along a diagonal, breaking four fold rotational symmetry. As the system magnetizes, these peaks become more diffuse and a Bragg peak centered in the first Brillouin zone emerges under large applied field indicating the emergence of long range ferromagnetic ordering.

Figure~\ref{fig:SSF_2D_averaged_pegasus_s0.3} displays a hysteresis loop and selected MSFs for a $\pm J$ model on a $32\times 32$ 2D square lattice, implemented on \texttt{Advantage\_system4.1} at an anneal fraction of $s=0.3$. The strong transverse field in the $\pm J$ model prevents the system from saturating, and under small longitudinal field the remnant magnetization disappears. This can be understood by examining selected MSFs. Around the coercive field (near zero longitudinal field), the MSFs have a combination of both ferromagnetic ordering and antiferromagnetic ordering, indicated with diffuse peaks at both the center and the middle of the edges of the first Brillouin zones. As the system demagnetizes and the remnant magnetization drops to nearly zero the ferromagnetic and antiferromagnetic peaks become of nearly equal intensity. As the system magnetizes under strong longitudinal field we can see the appearance of strong ferromagnetic ordering with strong Bragg peaks centered in the first Brillouin zone with otherwise diffuse intensity away from these Bragg peaks.

\begin{figure*}[th!]
     \centering
     \includegraphics[width=0.97\linewidth]{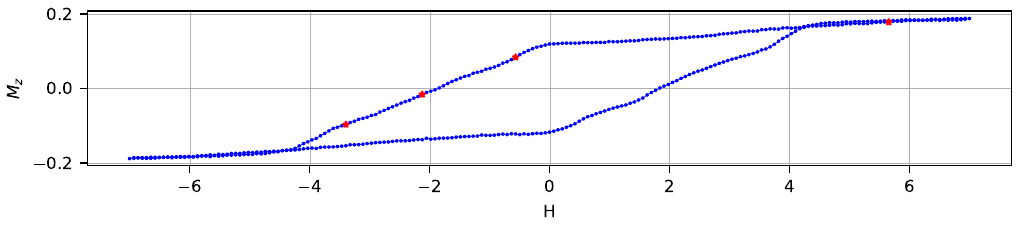}\\
     \includegraphics[width=0.246\linewidth]{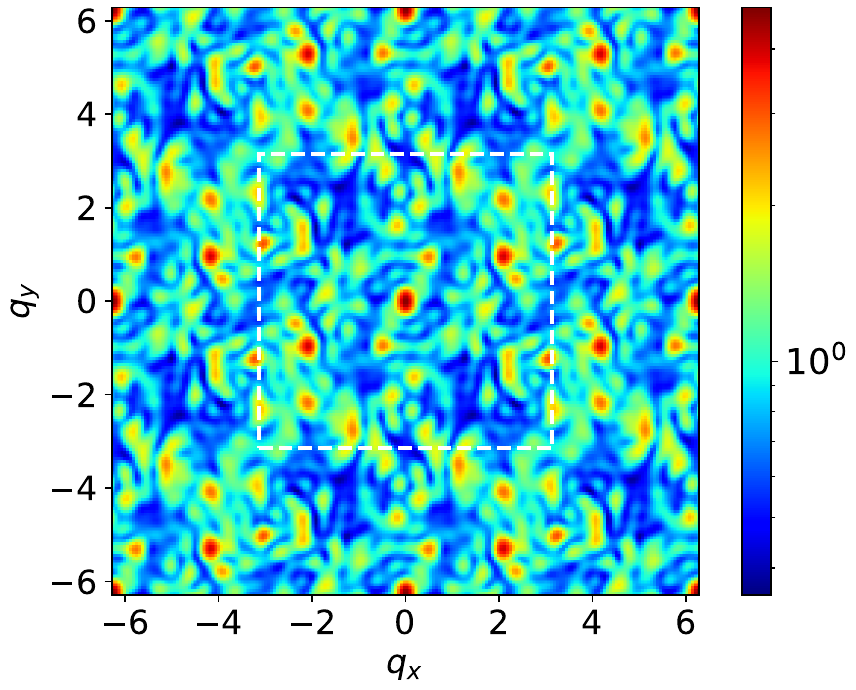}
     \includegraphics[width=0.246\linewidth]{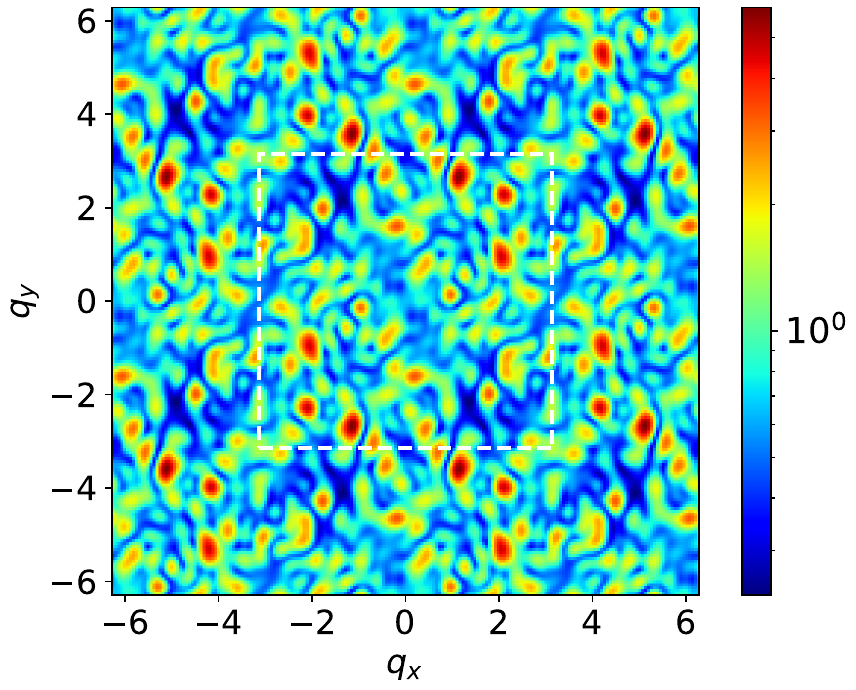}
     \includegraphics[width=0.246\linewidth]{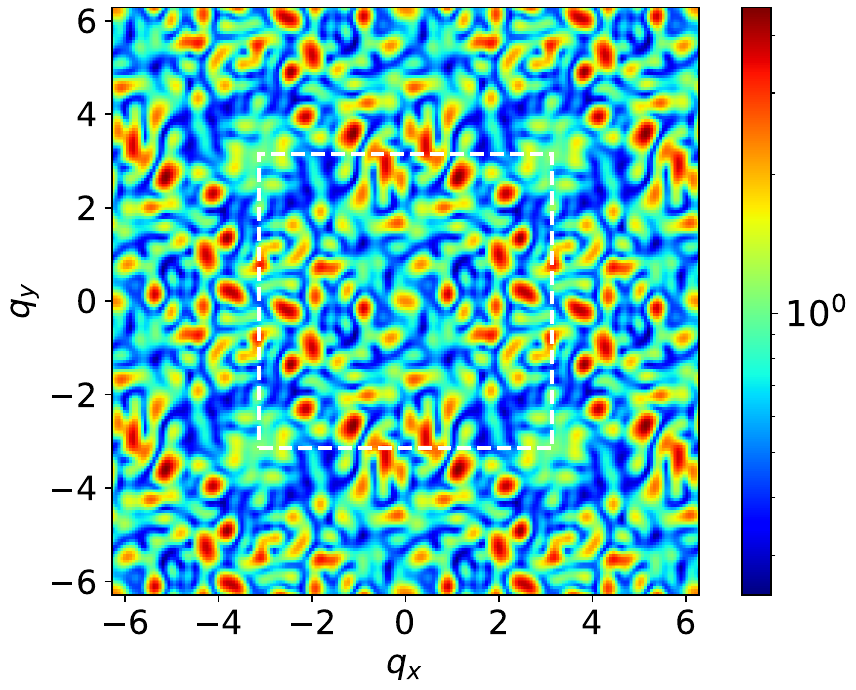}
     \includegraphics[width=0.246\linewidth]{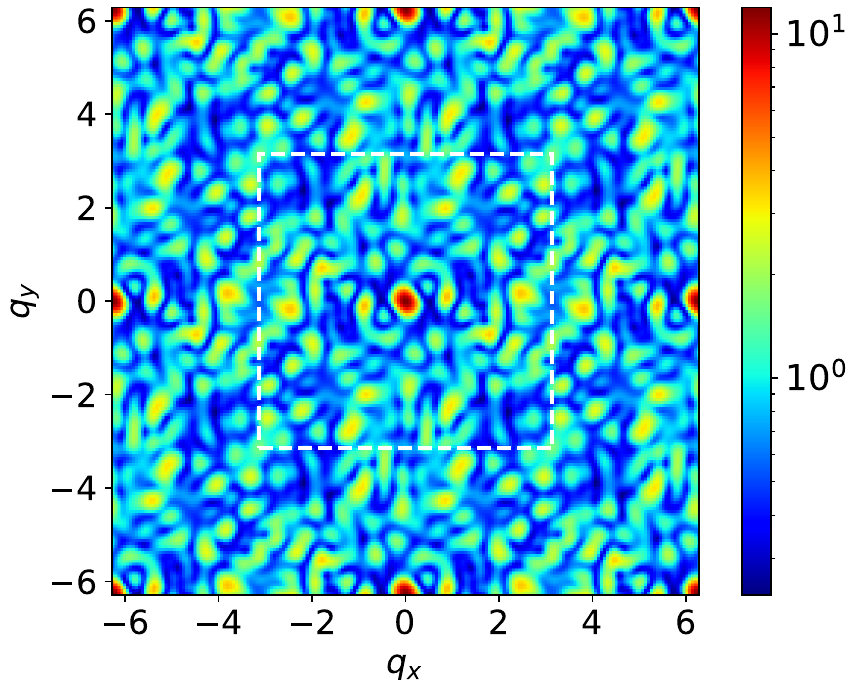}
     \caption{Averaged magnetic spin structure factor $\abs{S(q)}$, in log scale, from the quantum annealing hysteresis protocol at $s=0.3$ (stronger transverse field, and weaker $J$) from a $264$ node planar grid subgraph (e.g., 2D slice) of the Zephyr hardware graph \texttt{Advantage2\_prototype2.6} defined $\pm J$ model at four specific points along the hysteresis cycle. The hysteresis curve on the top plot is the average single site magnetization $M_z$ across the entire lattice, and the red asterisks denote specific points during the hysteresis cycle from which we have extracted averaged MSFs. The averaged MSF at each one of those points is given at the bottom - the order of these plots corresponds to the order of the red points on the hysteresis curve. Each MSF heatmap is averaged over the first $100$ spin configurations measured on the quantum annealing hardware for progressively longer-duration longitudinal field slices. Here, the four MSF plots do not share the same heatmap scale. The dashed white box outlines the first Brillouin zone. These hysteresis simulations achieved a smaller maximum magnetization at the maximum applied longitudinal field compared to the weaker transverse field simulations of Figure~\ref{fig:SSF_Zephyr_Slice_of_3D_spin_glass_averaged_s0.7}. }
     \label{fig:SSF_Zephyr_Slice_of_3D_spin_glass_averaged_s0.3}
\end{figure*}

\section{Magnetic Spin Structure Factor at a Non-Monotonic Dip}
\label{section:appendix_non_monotonic_dip_SSF}

Figure~\ref{fig:appendix_focused_example_non_monotonic_dip_SSF_plots} reports average magnetic spin structure factor plots at the three points (before, at, and after) the non-monotonic magnetization dip that was examined in Figure~\ref{fig:Figure_5_spin_structure_factor}-c. These three MSF heatmaps were capped at $\abs{S(q)}=30$ for improved visualization clarity.

The magnetization reversal shown in Figure~\ref{fig:appendix_focused_example_non_monotonic_dip_SSF_plots} is non-monotonic, resulting in a region of negative susceptibility shortly after the longitudinal field changes sign. Three corresponding spin structure factors are plotted, corresponding to the cyan, red and green points marked in the hysteresis loop. There is a train of sharp peaks along the primary axes of the magnetic spin structure factors due to the open boundary conditions of the square lattice. The correlation function $C(r)$ is obtained from the MSF principle axes and given in Figure~\ref{fig:Figure_5_spin_structure_factor}-c. In reciprocal space, the one dimensional cuts along $k_x=0$ and $k_y=0$ produce a series of peaks separated by $\Delta k =2\pi/32$, the envelope function of these peaks is the Fourier transform of $C(r)$. At the transient magnetization dip there is structural reordering. Before the dip there is a finite correlation length of $\xi\approx 1.7$, and the two-point correlation function $C(r)$ has an shoulder peak around $r=7$, reflecting domain-wall structure size in unit cells. At the dip, we find that $\xi\approx 0.8$, i.e., correlations decay faster than just before and after the dip, and residual correlations beyond dominant short range correlations are spread widely demonstrating domains span a wide distribution of scales. Beyond this point, the magnetic configuration coarsens, and correlations and domain structures similar to before the dip return. The transient fragmentation of domains is apparent in the real space spin configurations of Figure~\ref{fig:Figure_5_spin_structure_factor}-c. As the configurations are disordered, particularly at the dip, the peaks do not decay away from the central Bragg peak. When domains have a well defined characteristic scale the appearance of these structure peaks is suppressed. This transient negative susceptibility can be understood as a unique aspect of the TFIM and the $\pm J$ model, wherein when $H_z\approx 0$, the local environment, $h^\text{internal}$, experienced by each spin is reduced and the transverse field pushes the magnetization towards the paramagnetic regime. As the magnetization reversal continues, the system moves outside the regime dominated by the transverse field, the effective $Z$-basis Hamiltonian dominates again, and the still large-local effective field drives spins to align with the local magnetization, re-coarsening the magnetic configuration.

\section{2D Spin Configuration Plots}
\label{section:appendix_real_spin_configurations}

Since we have full measurements of the spin configurations, we can also visualize the spin values for specific samples. Figure~\ref{fig:real_config_2D_plots_4.1_s0.7} renders four single spin configurations, in the form of a pixel grid, from one of the $32\times 32$ 2D $\pm J$ models, at a single fixed longitudinal field value in an approximately de-magnetized region of the hysteresis sweep. These real space configuration plots are intended to be examples of the type of spin orientation and ordering that we see during this demagnetized portion of the hysteresis sweep. Examining Figure~\ref{fig:real_config_2D_plots_4.1_s0.7} we see a lack of ordering and consistent domain size. Thus the demagnetized configurations are not well ordered but instead disordered and demagnetized.

\begin{figure*}[ht!]
    \centering
    \includegraphics[width=0.32\linewidth]{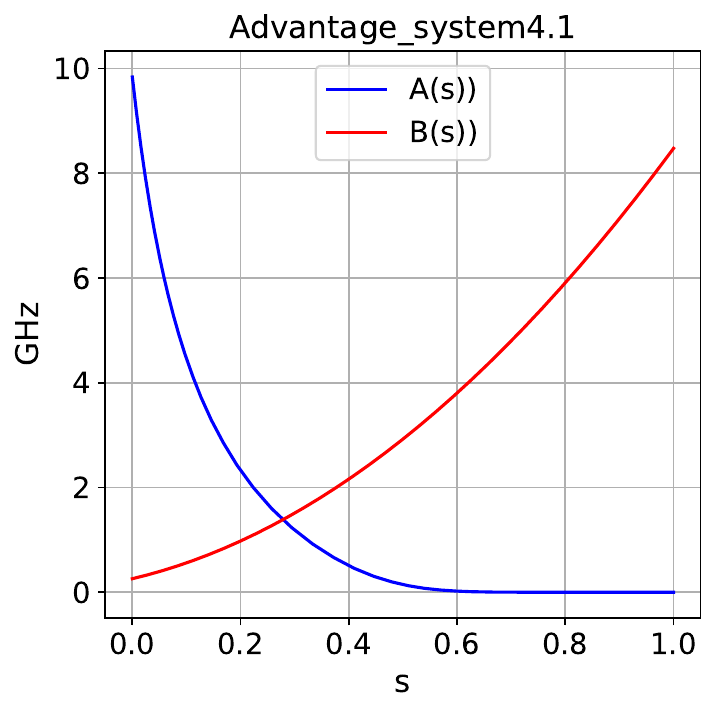}
    \includegraphics[width=0.32\linewidth]{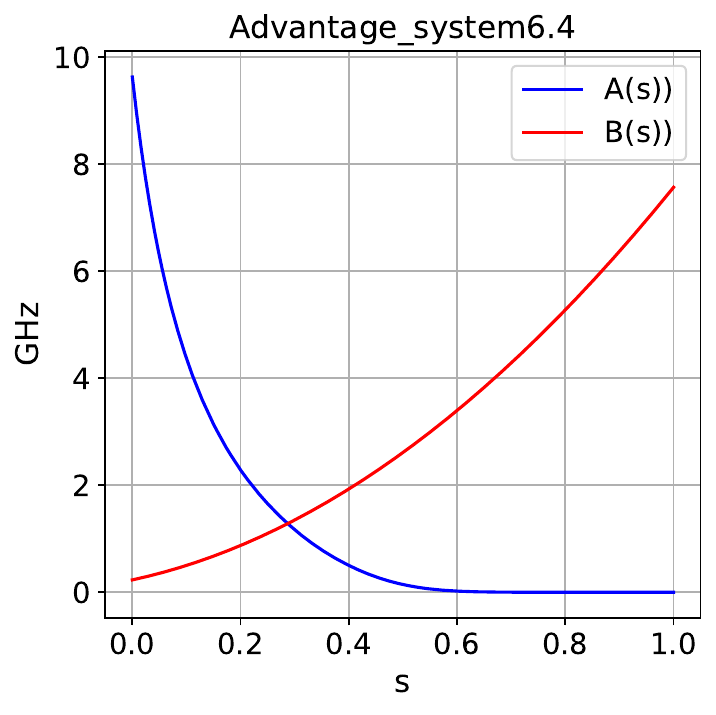}
    \includegraphics[width=0.32\linewidth]{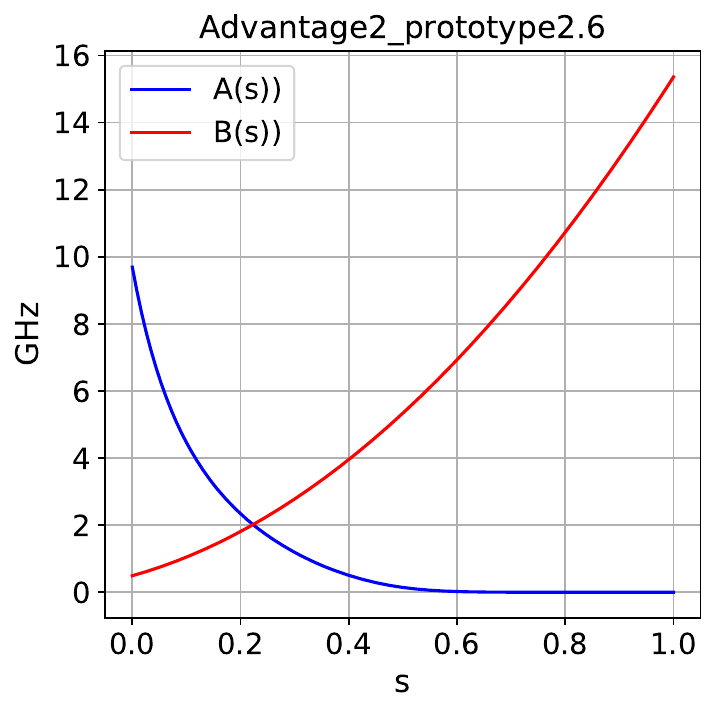}
    \caption{Energy scale calibrations for all three D-Wave QPU's used in this study. These exact quantities are shown for reference because these energy scales determine the physical analog simulation performed within the hysteresis protocol. }
    \label{fig:DWave_hardware_calibration_schedules}
\end{figure*}

\section{Strictly Planar 2D Square Grid Subgraph of the Zephyr Processor Graph for Magnetic Spin Structure Factor}
\label{appendix:planar_zephyr_subgraph}

Figure~\ref{fig:zephyr_graph_and_planar_subgraph} shows a fixed 2D planar square grid subgraph induced from the hardware graph of the \texttt{Advantage2\_prototype2.6} processor. This square grid can then be used to compute the MSF from a portion of the $\pm J$ model that is defined on the entire hardware graph. The key property that we need in order for this MSF to make sense is that the 2D slice of the somewhat densely interconnected hardware graph is strictly a planar subgraph meaning there no spurious interactions within this 2D grid, which is what Figure~\ref{fig:zephyr_graph_and_planar_subgraph} shows us. Note that this planar subgraph is somewhat arbitrarily chosen - there exist other strictly planar subgraphs of the Zephyr hardware graph, here we choose just one as a demonstration of extracting a 2D slice view of the spin correlations during the hysteresis cycle. Figure~\ref{fig:zephyr_graph_and_planar_subgraph} shows this subgraph. Figures~\ref{fig:SSF_Zephyr_Slice_of_3D_spin_glass_averaged_s0.7} and \ref{fig:SSF_Zephyr_Slice_of_3D_spin_glass_averaged_s0.3} show several averaged MSF heatmaps from this 2D slice of the full-hardware lattice defined $\pm J$ model on \texttt{Advantage2\_prototype2.6}, during different points of the hysteresis cycles. These slices being 2D planar subgraphs makes them valid MSF correlations, however, they are still components of a reasonably complex hardware graph and therefore difficult to visually interpret beyond a few clear observations. For computing these planar subgraph MSFs, we use the same techniques and sample averaging described in Supplementary Information~\ref{section:appendix_averaged_magnetic_spin_structure_factor}.

Figure~\ref{fig:SSF_Zephyr_Slice_of_3D_spin_glass_averaged_s0.7} displays a hysteresis loop along with four averaged MSFs along that magnetization curve for a $264$ node planar grid subgraph (e.g., 2D slice) of the $\pm J$ model defined on the \texttt{Advantage2\_prototype2.6} Zephyr hardware graph at $s=0.7$. We observe the system demagnetize through the disappearance of the strong Bragg peak centered in the first Brillouin zone (far left MSF) and the appearance of peaks around the first Brillouin zone, most apparent in the third MSF. The zig-zag pattern of diffuse intensity and bright peaks is believed to be a result of selecting the planar subgraph from the full hardware graph. The planar subgraph, shown in Fig.~\ref{fig:zephyr_graph_and_planar_subgraph} as red-dots in the full hardware graph are a planar graph in an otherwise higher-dimensional connectivity graph. The number of spins in the full-hardware graph between neighboring spins in the planar sub-graph is not a fixed number, nor is it consistently even or odd, thus we are not guaranteed of an ordered planar sub-graph even in an ordered full-hardware graph. Thus the zig-zag structure in the demagnetized MSFs indicates that the full-hardware graph is demagnetized and we may be capturing higher order structure, such as stripe domains or a N{\'e}el order of which the planar subgraph is a two-dimensional slice.

Figure~\ref{fig:SSF_Zephyr_Slice_of_3D_spin_glass_averaged_s0.3} displays hysteresis loop and example MSFs for the same $264$ node 2D slice of the $\pm J$ model defined on the \texttt{Advantage2\_prototype2.6} Zephyr hardware graph at $s=0.3$. Restricting our analysis of the magnetic ordering to the planar sub-graph we can see the MSFs corresponding to a partially demagnetized system have antiferromagnetic and ferromagnetic ordering. Interestingly, the antiferromagnetic ordering corresponds to stripe ordering that persists in the partially magnetized system throughout the magnetization reversal. This corresponds to the ordering of antiferromagnetic peaks which form diagonal stripes across the MSFs. These stripe domains are weak ordering as the MSF have persistent fluctuations and diffuse intensity indicating a lack of global order. Under strong field the system gains stronger ferromagnetic ordering, indicated by the stronger Bragg peaks in the center of the first Brillouin zones, away from these peaks there is diffuse intensity indicating fluctuating antiferromagnetic order, particularly noticeable are signal intensity around the corner of the first Brillouin zone indicating N{\'e}el ordering.

\section{D-Wave QPU $A(s)$ and $B(s)$ functions}
\label{appendix:DWave_calibrated_schedules}
Figure~\ref{fig:DWave_hardware_calibration_schedules} shows the exact energy scales of the D-Wave QPU control schedules $A(s)$ ($\Gamma$) and $B(s)$ (which denotes the energy scale of $J$). These are the quantities used in eq.\eqref{eq:H(s)_general}. The control parameter $s$ is critical for setting the simulation properties of the magnetic hysteresis protocol, but the exact quantities are device specific and in particular using these calibrated device schedules one can compute the ratio $\Gamma/J$.

\section{Numerical Data Analysis Details}
\label{appendix:numerical_sim_details}
The magnetic spin structure factors are computed with the help of the Python 3 library Numba \cite{10.1145/2833157.2833162} to speed up the computations, as well as Networkx \cite{SciPyProceedings_11} and Numpy \cite{harris2020array}. 

The area between the two sweeps of the magnetic hysteresis protocol are integrated numerically using two stages. First, is interpolation between all datapoints that are defined by averaged single-site magnetization $M_z$ on the hysteresis curve as a function of the applied longitudinal field. The interpolation used is the PCHIP 1-D monotonic cubic interpolation \cite{doi:10.1137/0905021} in scipy \cite{2020SciPy-NMeth}, using 10,000 points for the forward sweep and 10,000 points for the backward sweep. Then, the area between these curves is computed using the trapezoidal numerical integration rule in Numpy \cite{harris2020array}. All reported hysteresis areas are not normalized with respect the maximum possible area for each D-Wave device (meaning that the maximum possible areas for each device is different, determined by the hardware specifications).